\documentclass{article}[12pt]
\usepackage{amsmath,amssymb}

\begin{document}
\setcounter{page}{1} \pagestyle{plain} \vspace{1cm}
\begin{center}
\Large{\bf Dark Energy Density and IS(Israel-Stewart) Bulk Viscosity Model }\\
\small \vspace{1cm}
{\bf S. Davood Sadatian}$^{1,2}$, {\bf A. Saburi}$^{1}$\\
\vspace{0.5cm} {\it $^{1}$Research Department of astronomy $\&$
cosmology, University of Neyshabur, P. O. Box 9319774446, Neyshabur,
IRAN\\
$^{2}$ Department of Physics, Faculty of Basic Sciences, University
of Neyshabur , P. O. Box 9319774446, Neyshabur, Iran
 }\\email: {sd-sadatian@um.ac.ir}\\
\end{center}
\vspace{1.5cm}
\begin{abstract}
We investigate the thermodynamics of a dark energy bulk viscosity
model as a cosmic fluid. In this regard, the two theories of Eckart
and Israel-Stewart (IS) are the basis of our work. Therefore, we
first investigate the thermodynamics of cosmic fluids in the dark
energy bulk viscosity model and the general relationships. Then, we
express the thermodynamic relationships of Eckart's theory. Due to
the basic equations of Eckart's theory and Friedmann's equations, we
consider two states, one is $p=-\rho$ (standard) and the other is
$p\neq-\rho$ (non-standard). In the standard state, we define the
pressure $(p)$, energy density $(\rho)$ and bulk viscosity
coefficient $(\xi)$ of the cosmic fluid in terms of cosmic time and
we obtain its relations. We also mention that in this standard
state, because of $p=-\rho$, the value of $a(t)$ is zero, so $a(t)$
is not defined in this state. But in the non-standard case
$(p\neq-\rho)$ the bulk viscosity coefficient $(\xi)$ is zero and
only the scale factor and pressure and energy density of the cosmic
fluid is defined. We also consider two states of constant and
variable bulk viscosity coefficients and obtain three Hubble
constant parameters and scale factor in terms of cosmic time, and
energy density in terms of scale factor. In the state of variable
bulk viscosity coefficient, we consider the viscosity coefficient as
the power-law from energy density $(\xi=\alpha\rho^{s})$, which is
$\alpha>0$ and a constant. Following, we discuss about the
dissipative effects of cosmic fluids and examine the effects of
energy density for dark energy in the Israel-Stewart(IS) theory. The
results are comprehensively presented in two tables (1) and (2).
\\

{\bf Keywords}:{ Dark Energy, Viscosity, Effective Equation of
State.}\\

{\bf PACS}: 04.50.kd, 95.36.+x \\

\end{abstract}
\newpage

\section{Introduction}
Studies of the red shift of type Ia supernova in far galaxies
{[1,2]} showed that the rate of expansion of the universe not only
did not decrease but also increases, meaning that we are in an
accelerating universe. In addition, the expansion of the universe
can be studied through the field of cosmic microwave radiation {
[3]}.  One of the factors expressing the concept of the accelerating
universe is dark energy that contains more than 70\% of the total
energy of the universe and has a negative pressure. In fact the
universe is filled with a mysterious energy field whose effect on
the structure of large scales {[4]}. Dark energy uses in various
models to study the expansion of the universe including :
quintessence {[5,6]}, phantom {[7]}, K-essence {[8]}, chaplygin gas
{[9]}, modified gravity {[10,11]}, quantum model {[12]}, holographic
model {[13]}, dynamic models {[14]} and etc.  Given recent advances
in this area, however, the nature of dark energy is still
incomprehensible to us.  Another model that is considered for dark
energy is the perfect fluids model, this means, they consider the
dark energy of the universe as a perfect fluid in an isotropic world
that has a certain density, pressure, and temperature.  One of the
characteristics of fluids is the viscosity of fluids. In this
regard, there are important reviews on Dark Energy from viscous
fluids [15,16]. Authors in [15] presented the review of a number of
popular dark energy models, such as the $\Lambda$CDM model, Little
Rip and Pseudo-Rip scenarios, the phantom and quintessence
cosmologies with the four types (I, II, III and IV) of the
finite-time future singularities and non-singular universes filled
with dark energy. Also, Brevik et al considered the important
implications and the capabilities of the incorporation of viscosity,
which makes viscous cosmology a good candidate for the description
of Nature [16]. We know that viscosity divided into two types of
bulk and shear viscosity, shear viscosity is lost due to the
isotropic principle of the universe [17,18]. However, the perfect
fluid model for dissipative processes was first proposed by Eckart
{[19]}. Due to this theory was faced with limitations such as
non-causality and the propagation of unlimited dissipation
perturbations, Israel and Stewart proposed a generalized theory that
did not have these limitations {[20]} and this two theory
investigate the effects of bulk viscosity fluid as dark energy on
the expansion of the universe.  These kind of dark energy models
show that the accelerated expansion of the universe can be studied
without considering the cosmological constant {[21]}. In the bulk
viscosity model, it can be examined early and late in the cosmic
time.  The condition of accelerated expansion of the universe is a
violation of strong energy conditions, ie the sum of energy density
and three times the pressure of each component of the universe be
negative $(\sum_{i}{{\rho_{i}}+{3p_{i}}}< 0)$.  Due to this point
and also that dark matter does not emit radiation and is without
pressure, then $\rho_{i}/{3}>0$ in which case there must be a source
of negative pressure for the expansion of the universe which is
called dark energy. In the standard cosmological model, dark energy
is showed as a fluid with negative pressure and with the equation of
state $p=\omega\rho$ with a constant parameter $\omega$ that is
$\omega=-1$, so the corrections observed in the equation show that
$\omega$ can also change dynamically {[22]} and be $\omega< -1$
(crossing the phantom dividing line) {[23]}. If the parameter
 $\omega$ is assumed to be dynamic, it shows some thermodynamic
problems relate to the equation of state, such as positive entropy,
chemical potential and temperature {[24,25]}. One way to prevent
these thermodynamic problems is to assume dark energy as a perfect
fluid with bulk viscosity(actually called imperfect fluid).  Bulk
viscosity means that dissipative processes occur that violation the
predominant energy conditions, $p+\rho<0$ , and in this case dark
energy does not become phantom necessary {[26,27]}. In this case
there is a pressure called the effective pressure which is denoted
by $P_{eff}=p+\Pi$, where $p$ barotropic pressure and $\Pi$
viscosity pressure which is proposed in {[28]} by considering the
negative effective pressure in the bulk viscosity model of cosmic
fluids. Equilibrium fluids have no entropy and frictional heating
because they are reversible and without dissipative but real fluids
are irreversible. As mentioned earlier, dissipation processes in
perfect fluids were first described by Eckart, he modeled the
effective pressure of fluid as $\Pi= -3H\xi$ ($\xi$ is a function in
terms of cosmic time and energy density, and $H$ is Hubble's
parameter ). Crossing the phantom dividing line {[29]} , big rip
singularities using different values of the state equation parameter
($\omega$) in bulk viscosity {[30]} , dark fluid cosmology {[31]}
are among the applications of Eckart theory.  It is noteworthy that
most dark energy models are tuned to cosmic data and observations
and therefore do not require other experiments, so most dark energy
models can explain the expansion of the universe well. However, due
to the unknown nature of dark energy, dark energy models that could
be studied as cosmic fluid must be within the framework of known
laws of physics and thermodynamic laws. Several works on dark energy
thermodynamics have been described in {[18,32,33,34,35,36,37,38]}.\\
Therefore, in the first part of this paper, we express the
thermodynamics of cosmic fluids in general with a constant and
variable state parameter under the two theories of Eckart and
Israel-Stewart.  In the next section, we examine the dissipative
effects of perfect cosmic fluids and finally examine the effects of
energy density for dark energy in the
Israel - Stewart theory. \\

\section{Some aspects of Cosmic fluids in Eckart and Israel-Stewart Theories}
Let us first discuss a review of the cosmic fluids thermodynamics in
general. When we consider dark energy as a cosmic fluid, a state
equation is defined for it as
\begin{equation} p=\omega\rho \end{equation}

where $p$ is the cosmic fluid pressure and $\rho$ is the cosmic
fluid energy density considered for dark energy and $\omega$ is a
state equation parameter that can be both constant and variable
(relative to cosmic time or any other parameter).  On the other
hand, the Friedman equation for a homogeneous, isotropic and flat
universe based on the Friedman-Lematre-Robertson-Walker (FLRW)
parameter is defined as follows:
 \begin{equation} \dot{H}+ H^2=-\frac{1}{6}[\rho+3p] \end{equation}
where $H$ is the Hubble parameter and $ H=\frac{\dot{a}}{a}$ ,\
$H^2=\frac{1}{3}\rho$ and $a$ is the scale factor for such a
universe, also we consider the natural units $
 8\pi G = c =
 1$.  According to Friedman's equation, due to the effect of dark
energy density on the expansion of the universe, in the cosmic fluid
model, the equation of conservation of fluid energy density is
defined as
 \begin{equation} \dot{\rho}+3H(\rho+p)=0 \end{equation}
In the last two equations, the derivatives are in terms of cosmic
time.  If we want to consider the first law of thermodynamics in the
dark energy cosmic fluid model, we must first know that the internal
energy of this fluid is equal to
 $U=\rho V $, which $V$ is the physical volume of the fluid based on the time
scale factor, $V=V_0a^3$ is defined ( $V_0$ is the volume of the
fluid at the present time), so the first law of thermodynamics for
cosmic fluid will be equal to {[37]}:
 \begin{equation} TdS=dU+\rho dV-\mu dN \end{equation}
where $S$ is fluid entropy, $T$ fluid temperature, $U$ fluid
internal energy, $V$ physical volume of fluid, $\rho$ fluid energy
density, $\mu$ chemical potential and $N$ number of particles in the
fluid.

\subsection{Thermodynamics of Eckart theory}
Now we express the thermodynamics of cosmic fluids in Eckart theory.
Due to the investigation of the bulk viscosity of the cosmic fluid
in Eckart theory, the bulk viscosity pressure of the fluid under
Eckart theory is defined as follows [42]
 \begin{equation} \Pi=-3H\xi \end{equation}
where $\xi$ is the viscosity coefficient and can be a function in
terms of cosmic time or a constant. On the other hand, the equation
of conservation of cosmic fluids in Eckart's theory is equal to:
\begin{equation} \dot{\rho}+3H(\rho+P_{eff})=0 \end{equation}
where $ P_{eff}$ is called effective pressure which is used in
dissipative processes for bulk viscosity and is equal to:
\begin{equation} P_{eff}=p+\Pi \end{equation}
Where $p$ is called the barotropic equilibrium pressure, so
according to equations (5), (6) and (7) we have:
 \begin{equation}\dot{\rho}+3H(\rho+p)-9H^2\xi=0.\end{equation}
Equation (8) is the equation of conservation of cosmic fluids in
Eckart's theory that if we want to get another form of this
according to the parameter of the equation of state $\omega$ we
have:
\begin{equation} \dot{\rho}+3H(1+\omega)\rho-9H^2\xi=0. \end{equation}
Now we have to get an equation in terms of the Hubble parameter and
its derivatives (to determine the effects of dark energy and
Bulk-Eckart viscosity on the expansion of the universe), according
to Friedman's equation in relation (2) and $H^2=\frac{1}{3}\rho$ we
have:
 \begin{equation} 2\dot{H}+3H^2(1+\frac{p}{\rho})=3H\xi\end{equation}
Now, if we assume the state parameter $\omega$ to be constant and
solve the equation (10), the Hubble parameter obtain as a function
in terms of the bulk viscosity coefficient
\begin{equation} H(t)=\frac{e^{((\frac{3}{2})\int\xi(t)dt)}}{c+(\frac{3}{2})(p+\rho)\int e^{(\frac{3}{2})\int\xi(t)dt}dt} \end{equation}
where $c$ is an integral constant {[42]}.  According to $H
=\frac{{\dot{a}}}{a} $, If we integrate from equation (11),
 the scale factor obtain as follows:
\begin{equation}a(t)=A[(\frac{3}{2})(1+\frac{p}{\rho})\int e^{(\frac{3}{2})\int\xi(t)dt}dt+c]^\frac{2}{3(1+\frac{p}{\rho})} \end{equation}
where $A$ is another integral constant.  Note that this relation is
conditional $p\neq-\rho$, if $p=-\rho$, the scale factor is not
defined in the equations. If we consider $p=-\rho$ then we have
according to equation (8):
\begin{equation} \dot{\rho}=9H^2\xi \end{equation}
 On the other hand, in this case $(p = -\rho)$, we will have according to relation (10) $\omega=-1$.  As a result, the bulk
viscosity coefficient of the fluid will be equal to:
\begin{equation} \xi=\frac{2}{3} \frac{\dot{H}}{H} \end{equation}
 According to relation (13) and (14) we have:
\begin{equation} \dot{\rho}=6H\dot{H} \end{equation}
Now, by using equation (15), we will have:
 \begin{equation} \rho=3H^2+o \end{equation}
where $o$ is an integral constant, and if we assume its value to be
negligible, in other words, $H^2=\frac{\rho}{3}$, we arrive at the
Friedman equation, where the energy density of the fluid is called
the density of the dark energy. According to equations (1) and (16)
the equivalent fluid pressure will be equal to:
\begin{equation} p=3H^2\omega \end{equation}
So, if $p =-\rho$ is assumed, we obtain three equations $\xi(t)$,
$\rho(t)$, $p(t)$ for the cosmic fluid with bulk viscosity, and
$a(t)$ is not defined in this case. If $p\neq-\rho$, we obtain the
cosmic fluid as a cold dark matter in the standard model of
cosmology $(\Lambda CDM)$.  If the parameter of the equation of
state is unequal $-1$ $(\omega\neq -1)$ and $\xi=0$, according to
equation (12), the scale factor is rewritten:
\begin{equation} a(t)=a_0(1+\frac{3}{2}H_0(p+\rho)t)^\frac{2}{3(1+\frac{p}{\rho})} \end{equation}
 and the energy density is equal to:
 \begin{equation} \rho=\frac{3(H_0)^2}{(1+\frac{3}{2}H_0(1+\frac{p}{\rho})t)^2} \end{equation}
where $H_0>0$.  If $\xi(t)=\xi_0=const$ is assumed, the Hubble
parameter is equal to:
 \begin{equation} H(t)=\frac{H_0\xi_0e^{(\frac{3}{2})\xi_0 t}}{H_0(1+\frac{p}{\rho})(e^{(\frac{3}{2})\xi_0t}-1)+\xi_0}. \end{equation}
Now if we integrate from relation (20) (with condition $a_0 = 1$ )
then we have:
 \begin{equation} a(t)=a_0 [1+\frac{H_0}{\xi_0}(1+\frac{p}{\rho})(e^{(\frac{3}{2})\xi_0 t}-1)]^\frac{2}{3(1+\frac{p}{\rho})} \end{equation}
 In this case, the energy density will change
according to cosmic time as follows:
 \begin{equation} \rho(t)=\frac{3H_0^2 e^{3\xi_0 t}}{[(1+\frac{H_0}{\xi_0})(1+\frac{p}{\rho})(e^{(\frac{3}{2})\xi_0 t}-1)]^2} \end{equation}
Of course, energy density can also be obtained in terms of scale
factor, which in case we will have [42]:
\begin{equation} \rho(a)=3H_0^2[\frac{\xi_0}{H_0(1+\frac{p}{\rho})}+(1-\frac{\xi_0}{H_0(1+\frac{p}{\rho})})a^{\frac{-2}{3(1+\frac{p}{\rho})}}]^2. \end{equation}
Now if $\xi(t)=\xi(\rho(t))$, assuming different values for the bulk
viscosity coefficient, different results are obtained in the
equations, one of the hypotheses for the value of $\xi$ is
$\xi=\alpha\rho^s$ ($\alpha>0$ and $s$ is a constant), which is a
dependent viscosity coefficient to the power-law of energy density
{[39]}.  In this form, the viscosity coefficient in a particular
case, for example : if $\xi(\rho)=\rho^{\frac{1}{2}}$, creates a big
rip singularity at the late of the universe that applies only in
barotropic fluid with bulk viscosity {[42]}.  Now if $s=\frac{1}{2}$
is assumed and placed in relation (10) we will have:
 \begin{equation} 2\dot{H}+3H^2(1+\frac{p}{\rho}-\sqrt{3}a\rho)=0 \end{equation}
if we integrate from Equation (24), the Hubble function will be
equal to:
\begin{equation} H(t)=H_0[\frac{3}{2}H_0(1+\frac{p}{\rho}-\sqrt{3}a\rho)(t - t_0)]^{-1} \end{equation}
and if we integrate equation (25) again, the scale factor is
obtained:
 \begin{equation} a(t)=[1+\frac{3}{2}H_0(1+\frac{p}{\rho}-\sqrt{3}a\rho)(t - t_0)]^{\frac{2}{3(1+\frac{p}{\rho}-\sqrt{3}a\rho)}}. \end{equation}
With these relations (25) and (26) the energy density of the fluid
in the case of the variable value of the viscosity coefficient will
be equal to:
 \begin{equation} \rho(a)=3H_0^2(\frac{a}{a_0})^{-3(1+\frac{p}{\rho}-\sqrt{3}a\rho)} \end{equation}
with Assuming $a_0 = 1$, the last relation becomes
$3H_0^2a^{-3(1+\frac{p}{\rho}-\sqrt{3}a\rho)}$.  Because in this
case we assumed the viscosity coefficient to be variable, so with
the use of equation (14) we will have:
 \begin{equation} \xi(t)=\sqrt{3}aH_0[1+\frac{3(1+\frac{p}{\rho}-\sqrt{3}a\rho)H_0(t-t_0)}{2}]^{-1} \end{equation}
 In relation (27) in order for the energy density in terms of the scale factor to be able
to determine the expansion of the universe, the value of the state
parameter must be unequal $-1$ ($\omega\neq-1$)[42].  In other
words, $p\neq-\rho$, that in case the power of relation (27) remains
negative and the energy density and also the temperature decrease
and with decrease energy density , the universe is expanded.
Observations show that the value of $p+\rho$ is close to zero.  so,
to reduce the energy density, the condition $p+\rho>0$ and
$\sqrt{3}a\ll1$ is necessary.  In a model of dark energy is called
the phantom $\sqrt{3}a>p+\rho$, this means an increase in energy
density, which is completely opposite to the dark energy model of
cosmic expansion {[7]}.

\subsection{Thermodynamics of Israel-Stewart theory}
 The Israel-Stewart theory provides a better explanation than Eckart's theory and solves
the non-causal and instability problems of Eckart's theory.  In this
theory, we have a causal evolution equation for the bulk viscosity
pressure in the framework of Friedman equations, which is defined as
follows:
 \begin{equation} \tau\dot{\Pi} + \Pi=-3H\xi - \frac{1}{2}\tau\Pi[3H+\frac{\dot{\tau}}{\tau} - \frac{\dot{\xi}}{\xi}-\frac{\dot{T}}{T}] \end{equation}
where $\tau$ is the relaxation time for the bulk viscosity effects
and $T$ is the temperature change due to the bulk viscosity effects.
We see that the Israel-Stewart theory is more complex than the
Eckart theory.  If the temperature depends only on the energy
density and the density of the number of particles according to
{[27]} and {[37]} we have:
 \begin{equation} dT=(\frac{\partial T}{\partial\rho})_n  d\rho + (\frac{\partial T}{\partial n})_\rho  dn \end{equation}
 On the other hand, the equations of conservation
of energy density and particle density in the Israel-Stewart theory
are:
 \begin{equation} d\rho=-3H(\rho+p+\Pi) \end{equation}
 \begin{equation} dn=-3Hn \end{equation}
 Now if we replace relations (31) and (32) in relation (30) we
will have:
 \begin{equation} dT=-3H[(\frac{\partial T}{\partial\rho})_n  (\rho+p+\Pi)+(\frac{\partial T}{\partial n})_\rho  n] \end{equation}
The solution to this equation is $T=T_0\rho^\frac{p}{p+\rho}$.  This
solution actually shows a perfect fluid with a bulk viscosity in
Eckart's theory with condition that in equation (33) it is
$p\neq-\rho$ ($T_0$ is an approximate average value of temperature).
According to {[39]}, we consider a power- law for the bulk viscosity
coefficient which is
 \begin{equation} \xi=\xi_0\rho^s \end{equation}
where $s$ and $\xi_0$ are a constant and arbitrary parameter, also
$\xi_0>0$ .  Finally, according to {[42]} for relaxation time we
have:
 \begin{equation} \tau=\frac{\xi}{\rho}=\xi_0\rho^{s-1} \end{equation}
Now if relations (34) and (35) are the two main hypotheses for
solving equation (29), with consider Friedmann equations
(1),(2),(7), we'll have:
 \begin{equation} 2\dot{H}+ 3(1+\frac{p}{\rho})H^2 + \Pi=0 \end{equation}
Therefore, equation (36) can be substituted for the three equations
(33), (34), (35) and placed in the bulk viscosity pressure transfer
equation (29).  After placement, the differential equation can be
equated with the energy density conservation equation (31), which in
result the following relation is obtained as {[39]}:
\begin{equation}
\ddot{H}+ 3H\dot{H} + \frac{3^{1-s}}{\xi_0}\dot{H}H^{2(1-s)} -
(\frac{2p+\rho}{p+\rho})\frac{\dot{H}^2}{H}+\frac{9}{4}H^3(\frac{p}{\rho}-1)+
\frac{3^{2-s} (1+\frac{p}{\rho}) H^{2(2-s)}}{2\xi_0} =0
\end{equation}
Because the equation (37) is so complex, this equation can only be
modeled numerically. With giving the value of the parameter $s$, the
equation can be reduced to a simpler equation.  Note that for
$s\neq\frac{1}{2}$ and $s\leq-\frac{1}{2}$ there is no solution or
phantom model[42].  Israel-Stewart theory is a non-linear and causal
theory for dark energy cosmic fluid models, and a phantom solution
is needed to solve equation(37)(phantom solution correspond to
cosmic data) and also  for $ s\geq-\frac{1}{2} $ and $ s=\frac{1}{2}
$, there is a phantom solution for the equation(37) {[42,43]}.
Unlike the previous works done in various papers which was mostly
considered $ s=\frac{1}{2} $, we decided this time to use $ s=1 $
for consideration some aspects of model(means, the relaxation time
is constant). With placing $s=1$ in equation $(37)$ we will have:
 $$ \ddot{H}+ 3H\dot{H}+\frac{1}{\xi_0}\dot{H}-(\frac{2p+\rho}{p+\rho})\frac{\dot{H}^2}{H} +
 \frac{9}{4}H^3(\frac{p}{\rho}-1)+$$
 \begin{equation}\frac{3}{2}\frac{(1+\frac{p}{\rho})H^2}{2\xi_0} =0 \end{equation}
 with rewriting this equation we have:
$$ \xi_0\ddot{H}(1+\frac{p}{\rho})+
3\xi_0H\dot{H}(1+\frac{p}{\rho})+\dot{H}(1+\frac{p}{\rho})-
\xi_0(2p+\rho)\frac{\dot{H}^2}{H}+ $$
\begin{equation}\frac{9}{4}\xi_0 H^3(\frac{p}{\rho}-1)^{2}+
\frac{3}{4}H^2(1+\frac{p}{\rho})^2 =0
\end{equation} If $\xi_0=0$ then the bulk viscosity pressure will be
$\Pi=0$.  In this case, relation (39) will change as follows:
 \begin{equation}\dot{H}+ \frac{3}{4}H^2(1+\frac{p}{\rho})=0 \end{equation}
In equation (39), if $p=-\rho$ or $\omega=-1$ establishes the
standard cosmological state.  Therefore, if the bulk viscosity model
is presented as a solution to equation (39), it should be
$p\neq-\rho$ or $\omega\neq-1$, in other words, we should consider
the standard case as an exception.  One of the solutions of equation
(38) is Ansats, which is mentioned in {[27]} and {[40]} as follows:
 \begin{equation} H(t)=\frac{A}{t_\alpha-t} \end{equation}
 And in the general case in the form $\frac{A}{(t_\alpha - t)^p}$ which is $p<0$ and $t_\alpha$ is
time limited to the future which ends in a big rip singularity and
$A$ is a positive constant to describe the expansion of the
universe{[41]}.  If we assume the recent relation (41) as a solution
to the differential equation of relation (38) and also assume
$\xi_0=\xi(t)=t_\alpha -t$, the viscosity coefficient changes with
time until the universe reaches a future singularity time
$(t_\alpha)$.  An equation in terms of description constant of the
expansion of the universe ($A$) is obtained as follows:
 \begin{equation} \frac{9}{4}A^2(\frac{p}{\rho}-1)+ \frac{3}{4}A(\frac{p}{\rho}+5)+ \frac{p+2\rho}{p+\rho} =0 \end{equation}
 that the solution of this equation is:
 \begin{equation} A=\frac{-\frac{3}{4}(\frac{p+5\rho}{\rho})\pm \sqrt{{\frac{9}{16}}(\frac{p+5\rho}{\rho^2})-9(\frac{p-\rho}{\rho})(\frac{p+2\rho}{p+\rho})}}{\frac{9}{2}(\frac{p-\rho}{\rho})}. \end{equation}
Now to calculate the scale factor of this model, it is enough to
integrate the Hubble function relation (41), which in case we will
have:
 \begin{equation} a(t)=(\frac{t_\alpha - t_0}{t_\alpha - t})^A \end{equation}
where $t_0$ is the present time.  According to equations (32), (41)
and (44), we can calculate the density of the number of cosmic fluid
particles with bulk viscosity, which we have:
\begin{equation} n(t)=n_0(\frac{t_\alpha - t}{t_\alpha - t_0})^{3A} \end{equation}
From equation (44), it is obvious that if $t_\alpha = t$, the
expansion of the universe becomes infinite and the density of the
number of particles tends to zero, and thus through equation (41)
the bulk viscosity pressure can be obtained[41,42,43]:
 \begin{equation} \Pi(t)=A(2+3A(1+\frac{p}{\rho}))(t_\alpha - t)^2 \end{equation}
Note that the last relation establishes with condition
$H(t)=\frac{A}{t_\alpha - t}$ and can have different values.  In the
next section, using the thermodynamic relationships of this section,
we investigate the production of positive entropy of cosmic fluids
in dissipative processes and find that with increasing the entropy
of cosmic fluids with bulk viscosity as a model of dark energy, the
universe expands. \\

\section{Dissipative Effects of Perfect Cosmic Fluids}
In the study of the effects of bulk viscosity on cosmic fluids, as
mentioned earlier, the non-causal theory of Eckart and the causal
theory of Israel-Stewart (IS) have been defined. Bulk viscosity
models with dissipative processes can improve the expansion of the
universe calculation. In fact, entropy is produced in dissipative
processes lead to the expansion of the universe. Also entropy is
produced by isentropic particles according to IS theory {[44]}.  In
isentropic particles, the entropy of the particles is constant, but
due to the expansion of the universe, the entropy production of
these particles increases. It is noteworthy that if we consider
cosmic fluids openly so that the number of fluid particles is not
maintained, then the cosmic fluid becomes a non-equilibrium
thermodynamic system, and to create fluid particles in this state, a
bulk viscosity pressure,$\Pi$ , is defined.  In addition,
dissipative processes play a basic role in the evolution of the
early universe and also the big rip singularity.  In fact, they
violate the dominant energy conditions (DEC), which means that
$p+p+\Pi<0$, and on the other hand, because the bulk viscosity
pressure,$\Pi<0$, so these conditions increase the energy density of
fluid and according to equation (5), the previous bulk viscosity
coefficient is considered $\xi>0$. Therefore, with the two
conditions of violation of the dominant energy condition (DEC) and
 $\xi>0$, the entropy of the cosmic fluid increases and as a result, it
does not violate the second law of thermodynamics {[44]}.  Another
noteworthy point is that in the perfect causal theory of IS, in
order to obtain solutions that lead to big rip singularity, the dark
component must be considered as a phantom {[44]}. This is because
the dark energy of the phantom is inconsistent with the hypothesis
of a perfect cosmic fluid which the fluid is reversible and without
dissipative processes.  Therefore, if the dark component is
considered as a phantom then the cosmic fluids, like real fluids, is
irreversible and have dissipative processes in resulting the
production of entropy.\\
From the first law of thermodynamics $(4)$ also internal energy
relations, $U =V\rho$, volume $V=V_0a^{3}$ and the density of cosmic
fluid particles $n =\frac{N}{V}$, the following equation can be
written for cosmic fluid entropy with bulk viscosity pressure
$\Pi$:\begin{equation}nTdS=-3H\Pi\end{equation} If in the above
relation $\Pi<0$, the right side equation becomes positive and as a
result we conclude entropy production for the expanding universe.

\subsection{Thermodynamics of dissipative fluids}
First, before the entropy results, we check  the thermodynamics of
dissipated cosmic fluids.  Using the relation of the first law of
thermodynamics, ie $dE = dQ-PdV$ and relation $(4)$, we can write:
\begin{equation}TdS=dQ=dE+PdV\end{equation}
where $Q$ is the  exchanged heat and $E$ is the internal energy of
the dissipated fluid.  According to equation $(48)$, if no heat is
exchanged in the cosmic fluid, ie the universe is adiabatic $(dQ =
0)$, then we will not have entropy production $(dS = 0)$. Therefore,
one of the conditions for the expansion of the universe and the
production of entropy is that the universe  not be adiabatic.  If we
consider the two variables $V$ and $T$ as the two basic variables of
dissipative fluid thermodynamics, we get the criteria for $dS$
{[45]}:
\begin{equation}dLn(T)=dLn|1+\frac{p}{\rho}|-\frac{p}{\rho}dLn(V)\end{equation} Now if we consider
equation (49) with energy conservation equation $(3)$ we can write:
\begin{equation}dLn(T)=dLn|1+\frac{p}{\rho}|+dLn(\rho)+dLn(V)\end{equation}
 that If we integrate, we have:
 \begin{equation}\frac{V(2\rho+p)}{T}=\frac{V_0(2\rho_0+p_0)}{T_0}\end{equation}
On the other hand, the internal energy of cosmic fluids is defined
as $E=\rho c^{2}V$.  According to this equation and equation $(51)$,
the following equation can be obtained, which is called the modified
ideal gas law:
\begin{equation}\frac{E}{E_0}=\frac{T\rho}{T_0\rho_0}(\frac{p_0+\rho_0}{p+\rho})\end{equation}
 The dissipated cosmic fluid can be investigated under constant pressure and constant volume
conditions and the effects of their heat capacity ($C_p$ and $C_v$).
 The heat capacity of a cosmic fluid at constant pressure is defined
as follows {[44]}:
\begin{equation}C_p=(\frac{\partial h}{\partial T})_p\end{equation} that $h = E + PV$ is a fluid enthalpy.
Using equation $(52)$, $C_p$ can be written as follows:
\begin{equation}C_p=(1+\frac{p}{\rho})\frac{E}{T}=(1+\frac{p_0}{\rho_0})\frac{E_0}{T_0}=constant\end{equation}
and also for heat capacity at constant volume $C_v=(\frac{\partial
E}{\partial T})_v$ and using relations $(49)$,$(52)$ $C_p$ and $C_v$
are related to each other as
follows:\begin{equation}C_p=\frac{(1+\frac{p}{\rho})dLn(V)-dLn(\frac{p}{\rho})}{dLn(V)}C_v\end{equation}
 Now, if we consider $P$ and
$T$ as two independent thermodynamic variables of dissipated fluid,
the change in volume (expansion and compression) of the fluid can be
written as follows:
\begin{equation}\frac{dV}{V}=(\lambda dT-Ad\rho)\end{equation}
where $\lambda$ is the coefficient of thermal expansion at
constant pressure and $A$ is the coefficient of compression at
constant temperature and are defined as
follows:\begin{equation}\lambda=\frac{1}{V}(\frac{\partial
V}{\partial T})_p\end{equation},
\begin{equation}A=-\frac{1}{V}(\frac{\partial V}{\partial p})_T\end{equation}
In addition, another coefficient is defined for the adiabatic state
(constant entropy and temperature change) which is called the
compression coefficient in the adiabatic state and we
have:\begin{equation}B=-\frac{1}{V}(\frac{\partial V}{\partial
p})_S\end{equation} The noteworthy point of relations $(53)$ to
$(59)$ is that the ratio of heat capacities at constant pressure and
constant volume is equal to the ratio of compression coefficients at
constant temperature and constant entropy
{[47]}:\begin{equation}\frac{C_p}{C_v}=\frac{A}{B}\end{equation}
Also, in order for the thermodynamic system of the dissipated fluid
to be stable (depending on the phantom and non-phantom conditions
and the value range $\omega$), all coefficients $C_p$,$C_v$,$A$,$B$
must be positive {[44]}. Therefore, according to relation $(60)$ and
$(55)$, the relation between $A$ and $B$ can be written as follows:
\begin{equation}A=\frac{(1+\frac{p}{\rho})dLn(V)-dLn(\frac{p}{\rho})}{dLn(V)}B\end{equation} and also we have:
\begin{equation}\frac{\lambda}{A}=(\frac{\partial p}{\partial T})_v\end{equation}
\begin{equation}\frac{\lambda}{B}=(\frac{\partial p}{\partial T})_v\end{equation}
 Now, using the modified ideal gas
equation $(54)$ and equation $(49)$, the coefficient of thermal
expansion $\lambda$ can be written as follows:
\begin{equation}\lambda=\frac{C_p}{(1+\frac{p}{\rho})V}[1+\frac{\rho dp}{\frac{p}{\rho}[\rho dp-\frac{p}{\rho}(1+\frac{p}{\rho})dLnV]}]\end{equation}
 that from this relation we obtain the
following equations[44]:
\begin{equation}\lambda=\frac{C_p}{V}A\end{equation}
\begin{equation}\lambda=\frac{C_p^{2}}{VC_v}B\end{equation}
Here, we consider two different scenarios for cosmic fluids:\\
1. The cosmic fluid system is non-adiabatic.\\
2.  The cosmic fluid system be considered adiabatic. \\
If the fluid is non-adiabatic, ie entropy production is done, then
we consider  the temperature to be almost constant. In this case,
because entropy  production is done, so $B=0$. Also, the fluid has a
bulk viscosity  with negative pressure,$\Pi<0$, and its value is
greater than the equilibrium pressure, so the effective pressure
becomes negative $(p_{eff}<0)$ and as the pressure of the fluid
system decreases, the
 volume of the fluid system increases $(V>0)$( this is easily
 understood from equation $(58)$). On the other hand, according to the
 relations $(62)$ and $(65)$ in the constant volume process, with decreasing
 pressure, the temperature of the fluid increases.  Therefore, with
 the expansion of the cosmic fluid, the entropy and temperature both
 increase, and also according to equation $(65)$, the changes between
 the coefficients of expansion and compression of the cosmic fluid
 change linearly, as shown in Fig.1. This scenario is a acceptable conclusion from the standard theory
 of cosmology that with pass the time, the universe expands and its
 entropy and temperature increase.\\
\begin{figure}[h]
\begin{center}\includegraphics{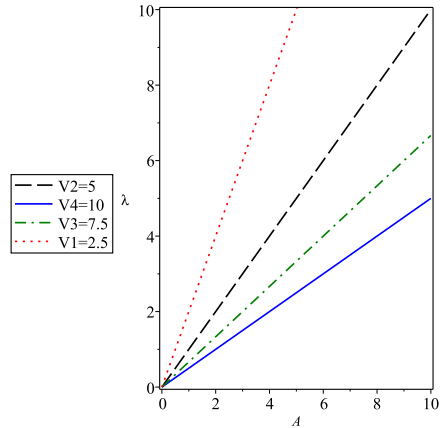} \vspace{6.5cm}
\end{center}
 \caption{\small {Ratio of expansion coefficient to
compression coefficient in non-adiabatic fluid. It is observed that
as time passes and the volume increases, the rate of expansion of
the universe decreases relative to its compression. The slope of the
diagrams is equal to $\frac{C_p}{V}$ with $C_p$ constant($C_p=5$)
and $V$ increasing ($V=2.5,5,7.5,10$).}} \label{Fig.1.}
\end{figure}

If the fluid is adiabatic, ie we do not have entropy production and
the entropy is assumed to be constant and unchanged, then we
consider $A=0$.  In such a cosmic fluid, the viscosity pressure of
the bulk is negative and the pressure decrease continuously in the
state of adiabatic expansion, and according to the equation $(59)$
the volume of the fluid increases, in this type of fluid(of course
in the adiabatic expansion), the temperature decrease with
decreasing pressure and increasing the volume of fluid. The
difference between this scenario and the previous scenario is that
first, we do not have any entropy production in this type of fluid,
but in the previous scenario, which is our acceptable scenario, the
entropy increases. The second difference is that the expansion rate
in this type of fluid (adiabatic) is more than Non-adiabatic fluid,
meaning that the universe expands more rapidly in this scenario and
also in previous scenario the temperature was increasing but in this
scenario the temperature decrease.
However, this scenario can have two specific situations:\\
Case 1: If the $C_v$ is constant and the $C_p$ increases then the
rate of expansion will be faster with pass the time(Fig. 2. a).\\
Case 2: If $C_v$ decreases and $C_p$ is constant: then the rate of
expansion increases for a limited time and then begins to compress
(Fig. 2. b). This case expresses the dark energy of the phantom
well, but is not acceptable in terms of standard cosmological
theory.  As we can see, the best model for the expansion of the
universe is Scenario 1, which is consistent with both cosmological
observations and standard cosmological theory.  This means that the
universe continues to expand, but its rate of expansion increases.

\begin{figure}[h]
\begin{center}\includegraphics{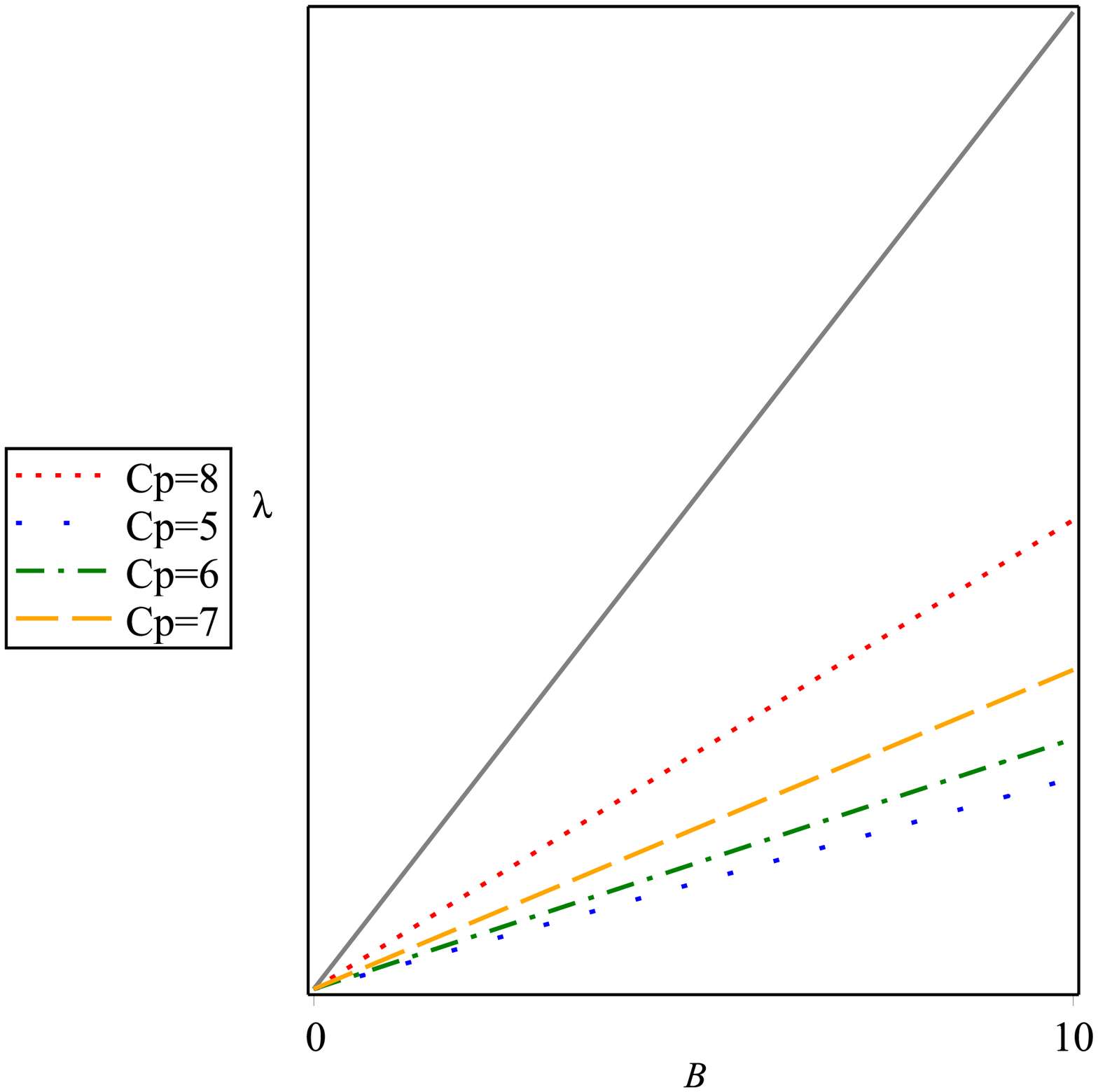} \vspace{12cm}\includegraphics{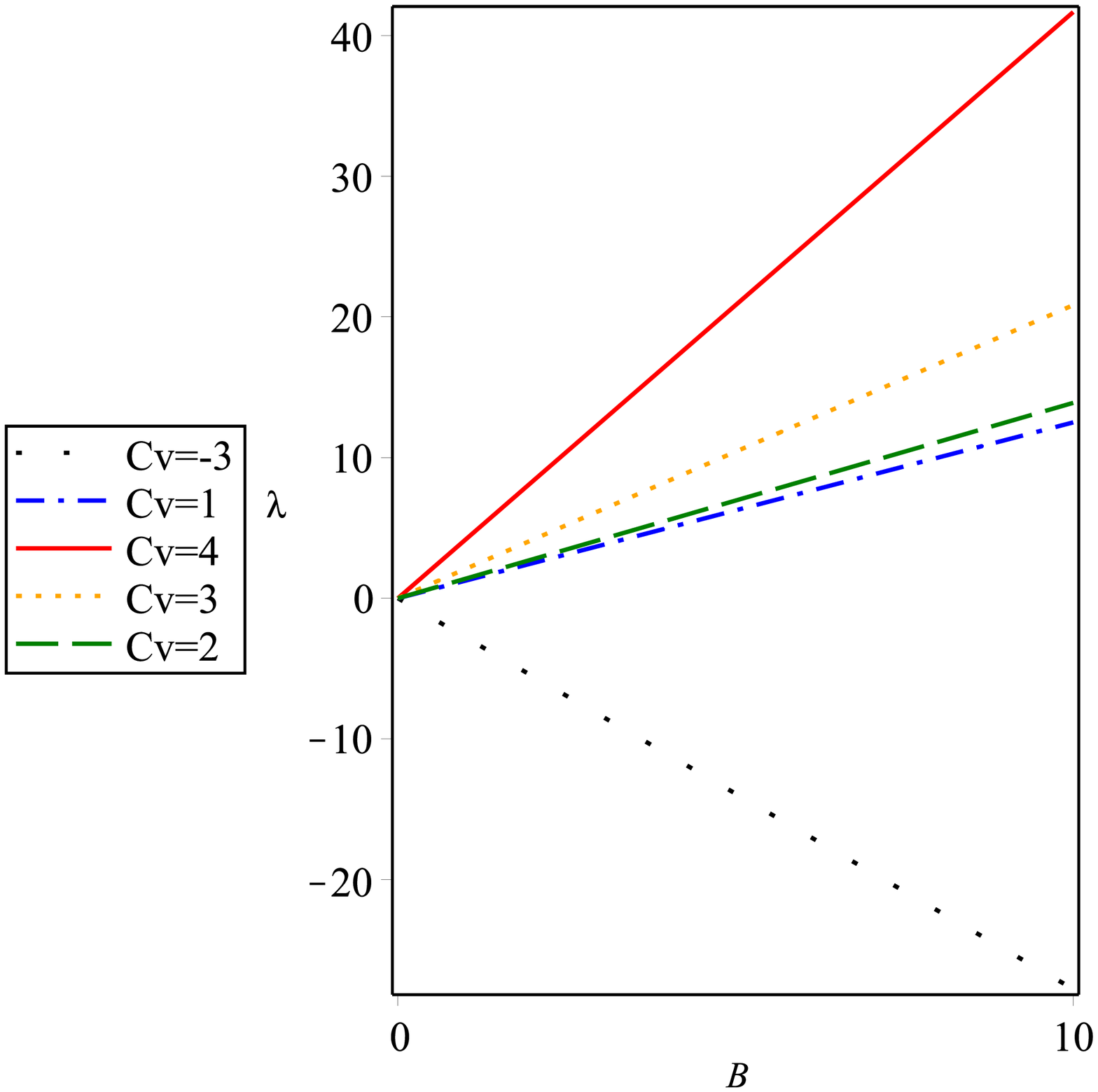}
\end{center}
\caption{\small {Fig.2.a(Up): The universe expand to infinity. The
slope of the  diagrams is equal to $\frac{C_p^2}{VC_v}$ that $C_v$
is constant($C_v=3$) and with increasing volume($V=1,3,6,10$) $C_p$
increase.($C_p=5,6,7,8)$. Fig.2.b.(Down): The universe expands until
a certain time and then stops accelerated expansion and starts to
compression.}}
\end{figure}

\subsection{Investigation of entropy of dissipative fluids}
First, we study the entropy of perfect fluids and then we will
investigate the entropy relations in the framework of the two
theories of Eckart and IS.  For a perfect fluid, in general terms,
two expressions of the equation of state, $p =\rho\omega$, with
$\omega$ constant and the particle flow number conservation
$n^{\alpha}=nu^{\alpha}$, where $u^{\alpha}$ is 4-velocity are
considered and we have:
\begin{equation}\dot{n}+3Hn=\frac{\dot{N}}{N}=0\end{equation}\\
 If we place equations $(67)$ and $(1)$ in the equations of conservation of
energy density $(3)$ and the first law of thermodynamics $(4)$,
respectively, the following relations are obtained:
\begin{equation}\dot{\rho}=-3H(p+\rho)\end{equation} \begin{equation}TdS=Vd\rho+(p+\rho) dV\end{equation}
that equation $(69)$ is also called the Gibbs relation. Now to
calculate the entropy with constant $\omega$ we have
\begin{equation}
TdS=(1+\omega)\rho
dV+Vd\rho=U(dLn(\rho)+3(1+\frac{p}{\rho})dLn(a))=0\end{equation}
This equation is obtained from equations $(68)$ and $(69)$.
Considering this relation, in this case, if $\omega$ is assumed to
be constant, then the entropy is constant, which means that the
fluid system is adiabatic, as we saw in scenario 2 in the previous
subsection.  Now to calculate the constant entropy value, we use the
following Eulerian relation:
\begin{equation}TS=U+\rho V-\mu N\end{equation}
Considering that the temperature and internal energy of a fluid are
thermodynamically proportional to each other $(U\propto T)$, hence
we can write the following
relation:\begin{equation}S=(1+\frac{p}{\rho})\frac{U}{T}-\frac{\mu
N}{T}=(1+\frac{p}{\rho})\frac{\rho V}{T}-\frac{\mu
N}{T}\end{equation} In perfect cosmic fluids with constant $\omega$,
entropy $S$ and the number of fluid particles $N$ are constant, so
according to equation (72) the $\frac{\mu}{T}$ ratio must be
constant, which $\mu$ is called the chemical potential of the fluid.
And based of the chemical potential, three states occur for the
cosmic fluid:\\ 1.  If $\mu=0$ in this case
$p\geq-\rho$ which in this case the cosmic fluid is far from the behavior of the phantom. \\
2.  If $\mu>0$ in this case $p>-\rho$.  The entropy has the least
constant value and the cosmic fluid in this case does not show the
behavior of the phantom. \\ 3.  If $\mu<0$ in this case $p<-\rho$.
We will have a phantom dark energy or
phantom cosmic fluid{[25]}. \\
In case $\omega$ is the variable $(\omega (a))$, if the chemical
potential is $\mu=0$, then the parameter of the variable state,
$\omega(a)$, is a constant parameter and is always greater than or
equal to $-1$, $\omega(a)\geq-1$, and if $\mu>0$ then $\omega(a)>-1$
and away from the standard state and if $\mu<0$ two states occur,
$|\mu|<\frac{S_0T}{V_0n_0}$ which in this case again $\omega(a)>-1$
and if $|\mu|>\frac{S_0T}{V_0n_0}$, we will have the phantom state
which in this case will be $\omega(a)<-1$ {[25]}. \\
Now in the framework of Eckart theory , we investigate the entropy
of dissipated cosmic fluids.  In this theory, we use equation $(47)$
to calculate the fluid entropy, which in instead of the bulk
viscosity pressure $-3H\xi$, therefore we will have:
\begin{equation}dS=\frac{9H^{2}\xi(t)}{nT}\end{equation}
Now we calculate the thermodynamic variables of density of particles
$n$ and temperature $T$ and place them in equation $(73)$, and after
we integrate from the obtained equation, finally we obtain the
entropy $S$ .  In here, we consider two cases:\\ 1.
$\xi(t)=\xi_0=constant$ and $\xi_0>0$ : according to the relations
$(3)$ and $(4)$ in the appendix and also the particle density
$n=\frac{n_0}{a^{3}}$, for $T=T_0\rho^{\frac{p_{eff}}{1+p_{eff}}}$
we have:
$$T(t)=T_0\rho_0^{\frac{p_{eff}}{\rho+p_{eff}}}\times$$
\begin{equation}[\frac{\xi_0e^{\frac{3}{2}\xi_0(t-t_0)}}{H_0(1+\frac{p_{eff}}{\rho})(e^{\frac{3}{2}\xi_0(t-t_0)}-1)+\xi_0}]^{\frac{2p_{eff}}{\rho+p_{eff}}}\end{equation}
\begin{equation}n(t)=n_0[1+\frac{H_0}{\xi_0}(1+\frac{p_{eff}}{\rho})(e^{\frac{3}{2}\xi_0(t-t_0)}-1)]^{\frac{-2\rho}{\rho+p_{eff}}}\end{equation}
 Now if we place two relations $(74)$ and $(75)$ in relation $(73)$
we have the following relation:
\begin{equation}\frac{dS}{dt}=\frac{3\xi_0\rho_0^{\frac{\rho}{\rho+p_{eff}}}}{n_0T_0}e^{\frac{3\xi_0\rho}{\rho+p_{eff}}(t-t_0)}\end{equation}
If we integrate from equation $(76)$ with assuming $p =-\rho$, the
entropy is
 obtained: \begin{equation}S(t)=S_0+\frac{(1+\frac{p_{eff}}{\rho})\rho_0^{\frac{\rho}{\rho+p_{eff}}}}{n_0T_0}(e^{\frac{3\xi_0\rho}{\rho+p_{eff}}(t-t_0)}-1)\end{equation}
where $S_0$ is a integral constant or the entropy at the present
time, now we can see, the increase in entropy is exponential. If in
the Eckart theory $\xi$ be variable and according to the hypotheses
of the previous section we consider $\xi=\alpha\rho^{s}$
$(\alpha>0)$, similar to the previous case, we get the density of
the number of particles $(n)$ and temperature $(T)$ and then we
place in $(73)$ and so integrate from the obtained relation to
obtain the entropy for the variable $\xi$.  Because in this case the
coefficient the viscosity of the bulk changes as power-law, choosing
a solution to obtain entropy will not be easy, but by giving a
specific value for $s$ and $\alpha$, a simpler solution can be
expressed. In a particular case, suppose
$\alpha=\frac{\sqrt{3}}{3}\gamma$ and $s=\frac{1}{2}$,
$\xi=(\frac{\sqrt{3}}{3}\gamma\rho^{\frac{1}{2}})$.  Now according
to equations $(6)$ and $(7)$ in the appendix to calculate the
density of the number of particles $n$, and temperature $T$, and
$n=n_0a^{-3},T=T_0\rho_0^{\frac{p_{eff}}{1+p_{eff}}}$ we have:
\begin{equation}n(t)=n_0[1+\frac{3H_0}{2}(1+\frac{p_{eff}}{\rho}-\gamma)(t-t_0)]^{\frac{-2\rho}{\rho+p_{eff}-\gamma}}\end{equation},
 \begin{equation}T(t)=T_0\rho_0^{\frac{p_{eff}}{1+p_{eff}}}[1+\frac{3H_0}{2}(1+\frac{p_{eff}}{\rho}-\gamma)(t-t_0)]^{\frac{-2p_{eff}}{1+p_{eff}}}\end{equation}
If we place the relations $(78, 79)$ in $(73)$, we obtain the
following differential equation:
\begin{equation}\frac{dS}{dt}=\frac{\gamma\rho_0^{\frac{2\rho+p_{eff}}{\rho+p_{eff}}}}{H_0n_0T_0}\times
[1+\frac{3H_0}{2}(1+\frac{p_{eff}}{\rho}-\gamma)(t-t_0)]^{\frac{-(\frac{p_{eff}}{\rho})^{2}+(2-\gamma)(\frac{p_{eff}}{\rho})+1}{(\frac{p_{eff}}{\rho})^{2}+(2-\gamma)(\frac{p_{eff}}{\rho})+(1-\gamma)}}\end{equation}
 We integrate from the above relation, therefore we have:
\begin{equation}S(t)=S_0+\frac{(1+\frac{p_{eff}}{\rho})\rho_0^{\frac{\rho}{\rho+p_{eff}}}}{n_0T_0}[(1+
\frac{3H_0}{2}(1+\frac{p_{eff}}{\rho}-\gamma)(t-t_0))^{\frac{\gamma-1}{(\frac{p_{eff}}{\rho})^{2}+(2-\gamma)(\frac{p_{eff}}{\rho})+(1-\gamma)}}-1]\end{equation}
In order for the entropy to increase in this case, the power of
equation $(81)$ must be positive. \\
To calculate the entropy in the Israel-Stuart (IS) theory, we have
the following differential equation that it obtain from the
placement Ansatz $(41)$ in equation $(46)$, density of the number of
fluid particles $(45)$ and the relation $(9)$ in the appendix
(temperature relation) in relation $(47)$, then we have:
\begin{equation}\frac{dS}{dt}=M(t_\alpha-t)^{N}\end{equation}
\begin{equation}M=\frac{(3A^{2})^{\frac{\rho}{\rho+p}}[2+3A(1+\frac{p}{\rho})](t_\alpha-t_0)^{3A}}{n_0T_0}\end{equation}
\begin{equation}N=\frac{2p}{\rho+p}-3(1+A)\end{equation}
Now we integrate from equation $(82)$ that we have(with condition
$p\neq\rho$):
\begin{equation}S(t)=S_0+(\frac{-M}{N+1})(t_\alpha-t)^{1+N}\end{equation}
where $S_0$ is an integral constant or entropy at the present time.
Here we present two scenarios:\\ 1.  If $M>0$ and $N<-1$ then we
conclude that $0<\frac{p}{\rho}<\frac{1}{2}$ that we shown in Fig.
3.a. With $\omega>0$ we conclude that the cosmic
fluid is expanding. \\
2. If $M<0$ and $N>-1$. in such a universe, entropy has a constant
value and then begins to decrease rapidly which totally violates
standard theory and Hubble's law(Fig.3.b).
\begin{figure}[htp]
\begin{center}\includegraphics{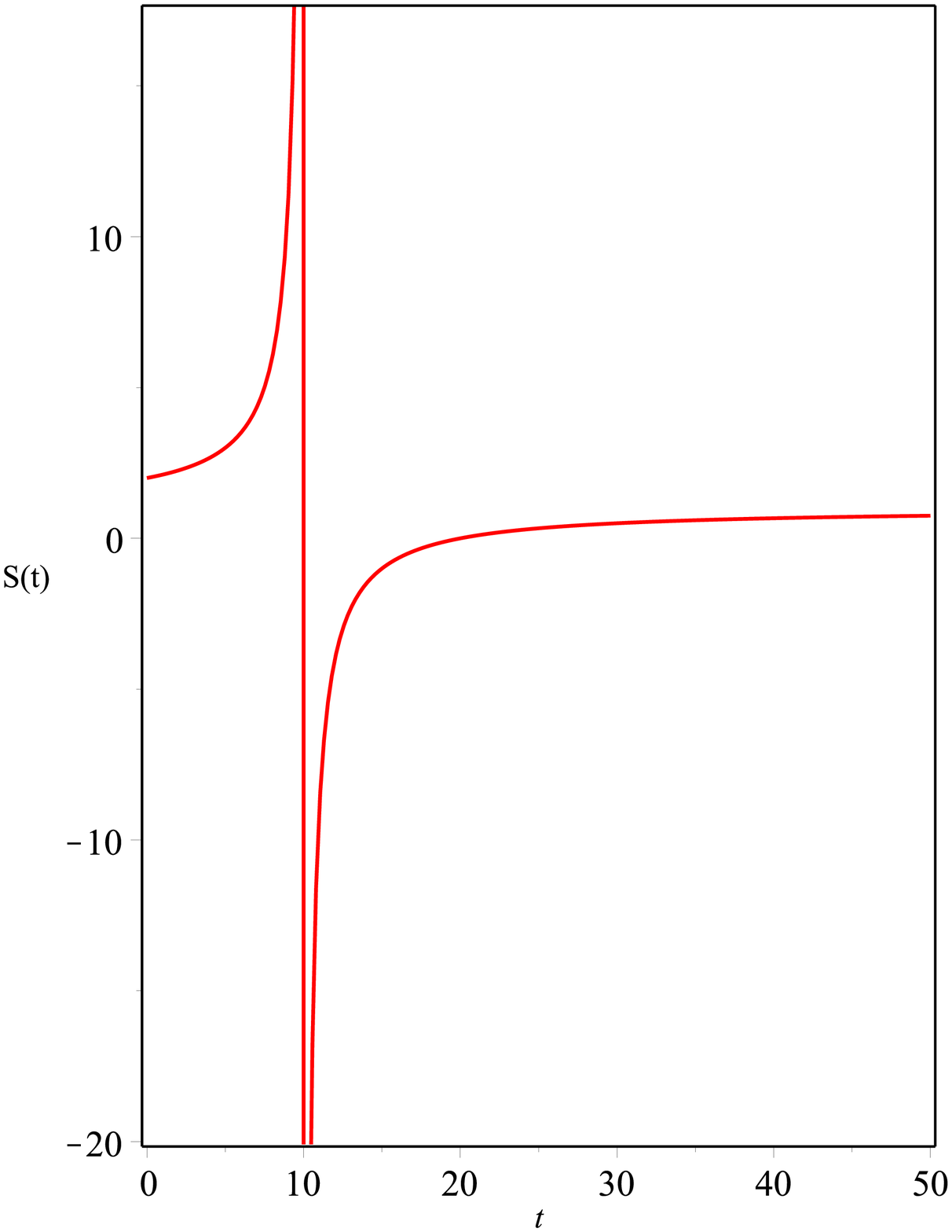} \vspace{12cm}\includegraphics{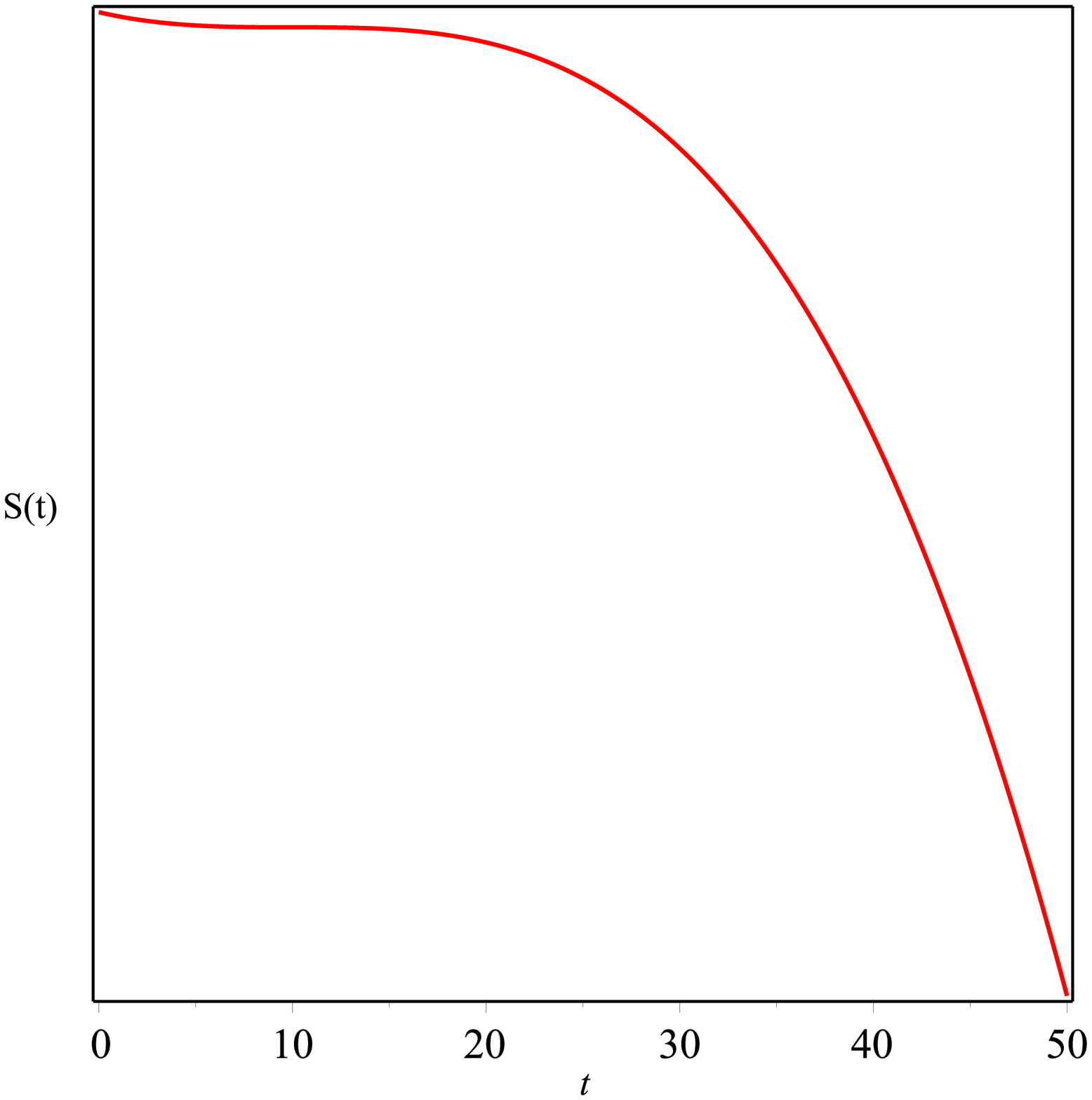}
\end{center}
\caption{\small {Fig.3.a(Up): Cosmic fluid entropy behavior over
time. We see that, the entropy increases infinitely up to a
asymptote(here assuming $t=10$ in term of Gpc). Then in the same
asymptote, the rate of expansion of the universe equals with its
rate of compression and after these asymptote, the average rate of
expansion of the universe be lower than the average rate of
compression until the average value of entropy production of
universe arrives a horizontal asymptote which means that the average
rate of expansion of the universe is closer to zero, that we
mentioned in the first scenario(for non-adiabatic fluid). with
hypothetical conditions[$M=10,N=-2,t_\alpha=10,S_0=1$].
Fig.3.b.(Down): In this state that it does not correspond to the
standard theory at all, the entropy is constant in a limited time
and then begins to decrease very rapidly which in fact shows an
unreal universe that contradicts Hubble's law. With hypothetical
conditions[$M=-10,N=2,t_\alpha=10,S_0=1$].}}
\end{figure}
\section{The Effects of Energy density for Dark Energy in the IS theory}
In this section, we want to investigate the effects of the energy
density $\rho$ in the framework of the Israel-Stewart theory. The
main parameters that its effects are investigated : temperature
$(T)$, bulk viscosity coefficient $(\xi)$ and relaxation time
$(\tau)$ (according to equations (29) to (35)). But each of these
parameters is obtained in terms of energy density $(\rho)$, so its
effects and relationships can be obtained. In this regard, if we
rewrite equation $(30)$ as energy density, we have:
\begin{equation}
d\rho=(\frac{\partial\rho}{\partial T})_{n}dT +
(\frac{\partial\rho}{\partial n})_{T}dn
\end{equation}
We place the density relation of the number of fluid particles
$(32)$ and the temperature differential equation $(33)$ in equation
$(86)$, we have:
\begin{equation}
d\rho=-3H[(p+\rho+\Pi)+ n(\frac{\partial\rho}{\partial T})_{n}
(\frac{\partial T}{\partial n})_{\rho}-
n(\frac{\partial\rho}{\partial n})_{T}]
\end{equation}
that we simplify this equation, we get the equation of conservation
of energy density in equation $(31)$:
$$d\rho=-3H(p+\rho+\Pi)$$
Due to the bulk viscosity pressure in the Israel-Stewart (IS)
theory, we have two states:\\
{\bf Mode 1:} $\Pi=-2\dot{H}-3(1+\frac{p}{\rho})H^{2}$ (according to
the equation $(36)$), we place this value $\Pi$ in the energy
density conservation equation $d\rho$ we have:
\begin{equation}
d\rho=-3H(p+\rho-2\dot{H}-3H^2(1+\frac{p}{\rho}))
\end{equation}
Now we integrate from the above relation twice, then we have:
\begin{equation}
\rho(a)=\frac{9}{20}(1+\frac{p}{\rho})H^{5}+(-\frac{1}{2}(p+\rho)+1)H^{3}
\end{equation}
In following, we summarize the results according to table $(1)$.
Also we plotted the obtained results in figure $(4)$. According to
figure $(4A)$ that shows the energy density in term of scale factor
in standard approach. The condition of the universe expansion and
increase of the scale factor is the ascending increase of the value
of energy density. In this approach, the pressure value is
symmetrical with energy density value and with the energy density
value increase, the value of pressure becomes more negative. In
figure $(4B)$ that shows the energy density in term of scale factor
in dynamical and non-phantom approach. In this figure pressure value
is greater than the energy density value, with the universe
expansion and the scale factor increase, the energy density value
increases similar in the standard approach. But in this case as much
as the energy density value be lower, the universe expands faster.
Finally, figure $(4C)$ shows the energy density in the phantom
approach, with the universe expansion and the scale factor increase,
the energy density value increases but in this case as much as the
energy density value be more, the universe expand faster and also if
the energy density value exceeds a certain value, the universe stops
accelerated expansion and begins to slow expansion and when energy
density value becomes infinite, the
universe will reach a big rip singularity in the future.\\

\begin{figure}[htp]
\begin{center}\includegraphics{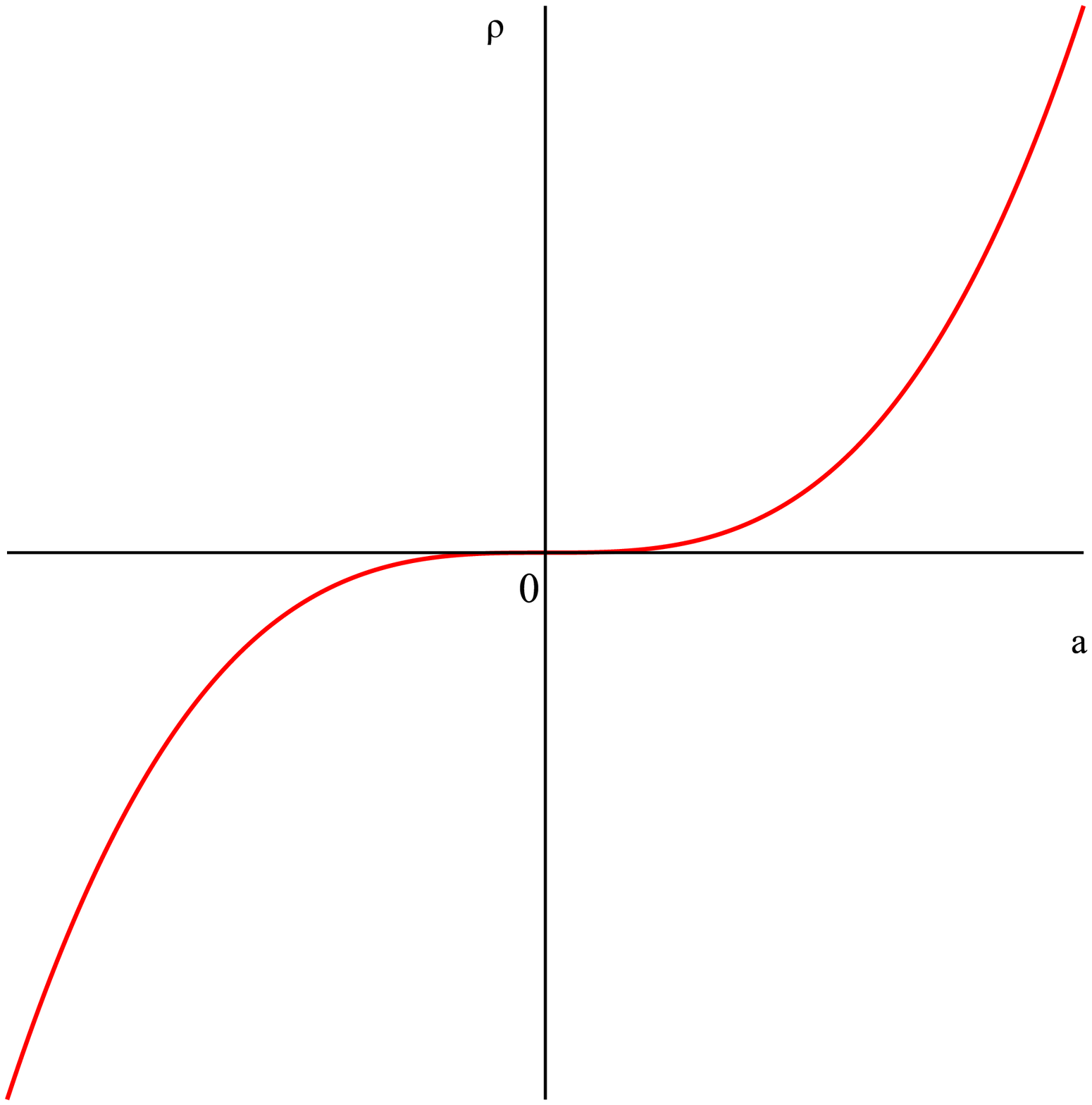} \vspace{8cm}\includegraphics{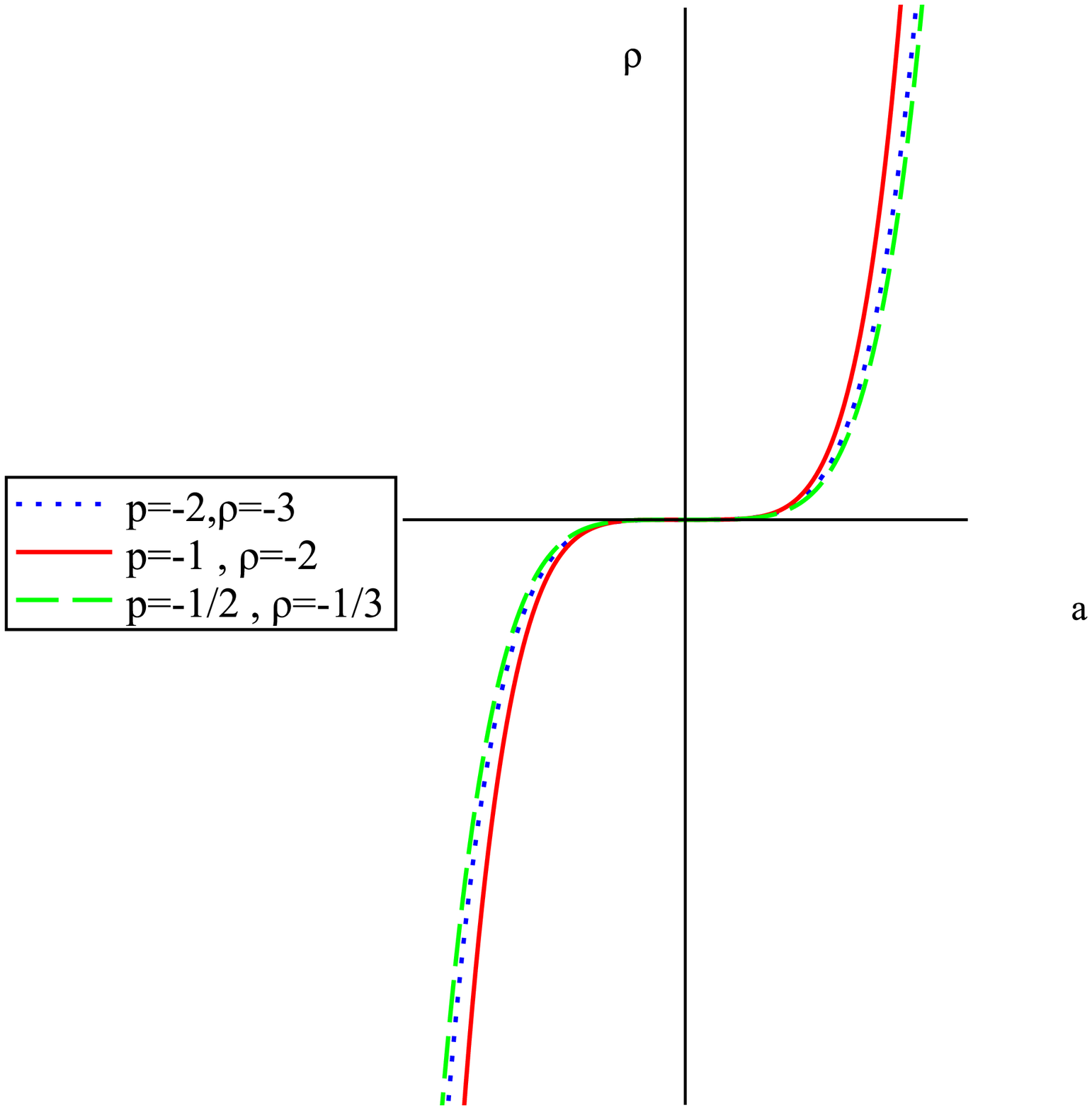}
\vspace{10cm}\includegraphics{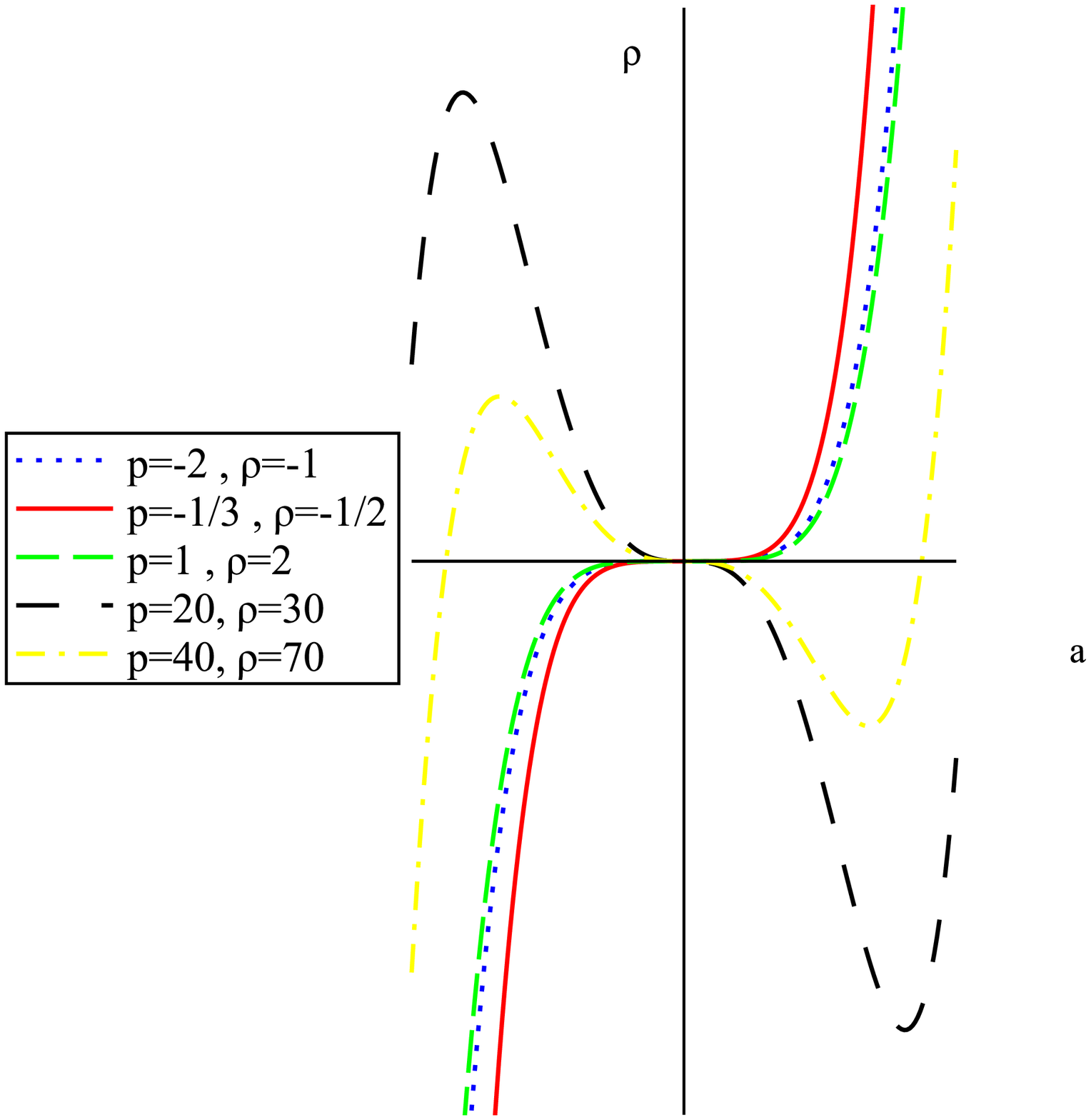}
\end{center}
\caption{\small {The effects of energy density in term of scale
factor on the universe expansion in Israel-Stewart theory(IS)
Fig.4.A(up): standard approach with condition $p=-\rho$,
Fig.4.B(middle): dynamical and non-phantom approach with condition
$p>-\rho$, Fig.4.C(down): phantom approach with condition
$p<-\rho$}.}
\end{figure}

{\bf Mode $2$:} $\Pi(t)=A(2+3A(1+\frac{p}{\rho}))(t_{\alpha}-t)^2$
(according to the equation $(46)$), we place this relation in the
energy density conservation equation  $d\rho$ so we have:
\begin{equation}
d\rho=-3H[p+\rho+A(2+3A(1+\frac{p}{\rho}))(t_{\alpha}-t)^2]
\end{equation}
And, as we know, $t_{\alpha}$ is a time that leads to a singularity
in the future, and $A$ is a constant that describes the expansion of
the universe(its value was obtained in equation $(43)$). Now if we
integrate from the above relation we have:
\begin{equation}
\rho(t)=-3H[pt+\rho
t-\frac{1}{3}A(2+3A(1+\frac{p}{\rho}))(t_{\alpha}-t)^3]
\end{equation}
\begin{table}[p]
\begin{center}
\caption{The results of the energy density effects in terms of scale
factor on the universe expansion in the framework of IS theory ({\bf
Mode 1}).} \vspace{0.5 cm}
\begin{tabular}{|c|p{4cm}|p{7.7cm}|}
  \hline
  \hline {\tiny Condition} & {\tiny Approach} & {\tiny Description} \\
  \hline {\tiny $P=-\rho$}& {\tiny Standard }& {\tiny In this case, the parameter of the cosmic fluid state equation ($\omega$) is considered constant as dark energy with bulk viscosity and its value is $\omega=-1$ and according to equation (1) it becomes $p=-\rho$ and also has no dynamics. In this type of approach the pressure value of cosmic fluid is negative and energy density increases and finally, the universe continues the expansion
  rapidly.}
  \\
  \hline {\tiny $P>-\rho$}& {\tiny Dynamical and non-phantom }& {\tiny In this approach, first the parameter $\omega$ is not constant and increase with the expansion of the universe, that is, as the universe expands, the pressure of the cosmic fluid with bulk viscosity tends from negative to positive value. This means that the energy density value decreases. Second, this approach conforms to the thermodynamics laws and the standard cosmological model and also shows that as the universe expands, the entropy and temperature increase and the chemical potential becomes
  negative [24],[25].}\\
  \hline {\tiny $P<-\rho$} &{\tiny Phantom} &{\tiny The energy density of a cosmic fluid increases as power-law($\rho=\rho_0 a^r$), that r is a constant and positive, and $\rho_0$ is the initial value of the energy density of the fluid. In this approach, with the growth of the scale factor and the expansion of the universe, the energy density increases and leads to  decrease in temperature and no entropy is produced. The rate of expansion in this approach is much higher and leads to a big rip singularity. This approach also violates the dominant energy conditions (DEC) and faces with many challenges and are classically and quantum
  unstable.[14]}\\
       \hline
\end{tabular}
\end{center}
\end{table}
We obtain the following results according to table $(2)$, we also
plotted the obtained results and relations in figure $(5)$.\\
In this mode, and figure $(5A)$ the results are similar to the
standard approach in table$(1)$, but with the difference that in
previous mode the energy density was in terms of scale factor but
here in terms of cosmic time and an expansion coefficient $A$ that
obtained in the second section. In figure $(5B)$ with over time and
the universe expansion, the energy density value decreases so the
pressure value increases and tends to be positive. Finally, in
figure $(5C)$ with over time, the energy density value increases and
as a result the universe is expanded faster but when it crosses the
phantom divide line (here the phantom divide line has been showed by
a black dotted line on the $\rho$ axis) and the energy density
becomes infinity, the universe begins to slow expansion until it
finally reaches a big rip singularity at the time $t_{\alpha}$. In
this section we could to study the effects of energy density in
terms of scale factor and in terms of cosmic time on the expansion
of the universe under the Israel-Stewart (IS) theory.\\
Here, Let us compared results of our research with earlier studies
of Dark Energy from viscous or inhomogeneous fluids according
references [48,49,50]. In this regard, Brevik et al considered the
role of a viscous (or inhomogeneous (imperfect) equation of state)
fluid in a Little Rip cosmology. Despite the earlier observations
that viscosity basically supports the Big Rip singularity, they have
demonstrated that it is also able to give rise to a non-singular,
Little Rip cosmology, which is considered to be a viable alternative
to $\Lambda$CDM cosmology. In particular, constant bulk viscosity
and a viscosity inversely proportional to the Hubble rate have been
considered. They have shown that in those cases a Little Rip
cosmology may naturally emerge. It is remarkable that for a standard
fluid, $p = \omega\rho$, the only influence of viscous effects can
naturally drive the universe to a Little Rip type evolution[48].
Also, Brevik and Timoshkin studied a general equation of state for
the dark fluid in the presence of a bulk viscosity. They have
explored the holographic principle for cosmological models with
various values for the thermodynamic parameter $\omega(\rho, t)$ and
for different forms of the bulk viscosity $\zeta(H, t)$. For each
model the infrared radius, in the form of a particle horizon, has
been calculated in order to obtain the energy conservation law.
Thus, they have shown the equivalence between viscous models and the
holographic model[49]. Finally, Nojiri et al in [50] considered the
effect of modification of general equation of state of dark energy
ideal fluid by the insertion of inhomogeneous, Hubble parameter
dependent term in the late-time universe. Several explicit examples
of such term which motivated by time-dependent bulk viscosity or
deviations from general relativity studied. The corresponding
late-time FRW cosmology (mainly, in its phantom epoch) described.
Also, they found that the inhomogeneous term in equation of state
helps to realize such a transition in a more natural way. It is
interesting that in the case when universe is filled with two
interacting fluids (for instance, dark energy and dark matter) the
Hubble parameter dependent term may
effectively absorb the coupling between the fluids [50]. \\
In comparing of above results with our findings in section 4, we can
point out in summary : 1- In standard approach, with over time, the
universe has an accelerated expansion (of course, before time
$t_\alpha$). After time $t_\alpha$, we assume that over time
pressure tends to $p\rightarrow -2\rho$ and the universe expand
slowly. In this approach the energy density value increases rapidly
but its value is symmetrical with the value of negative pressure. 2-
In dynamical and non-phantom approach, with over time, the pressure
value is negative and will change with the energy density value, and
expansion will increase, and as a result, the energy density will
decrease. Therefore, the condition for the expansion of the universe
with over time is the decrease in the dark energy density variable
value. As much as the energy density value is lower, the universe
expand faster. 3- In phantom approach, with over time, the energy
density becomes bigger and tends to be infinity. As a result, the
universe expands more rapidly, but after crossing the phantom divide
line, the value of energy density becomes infinity(the maximum value
of energy density), the universe’s expansion rate gradually
decreases and finally reaches a big rip singularity after time
$t_\alpha$ [51].\\
\begin{figure}
\begin{center}\includegraphics{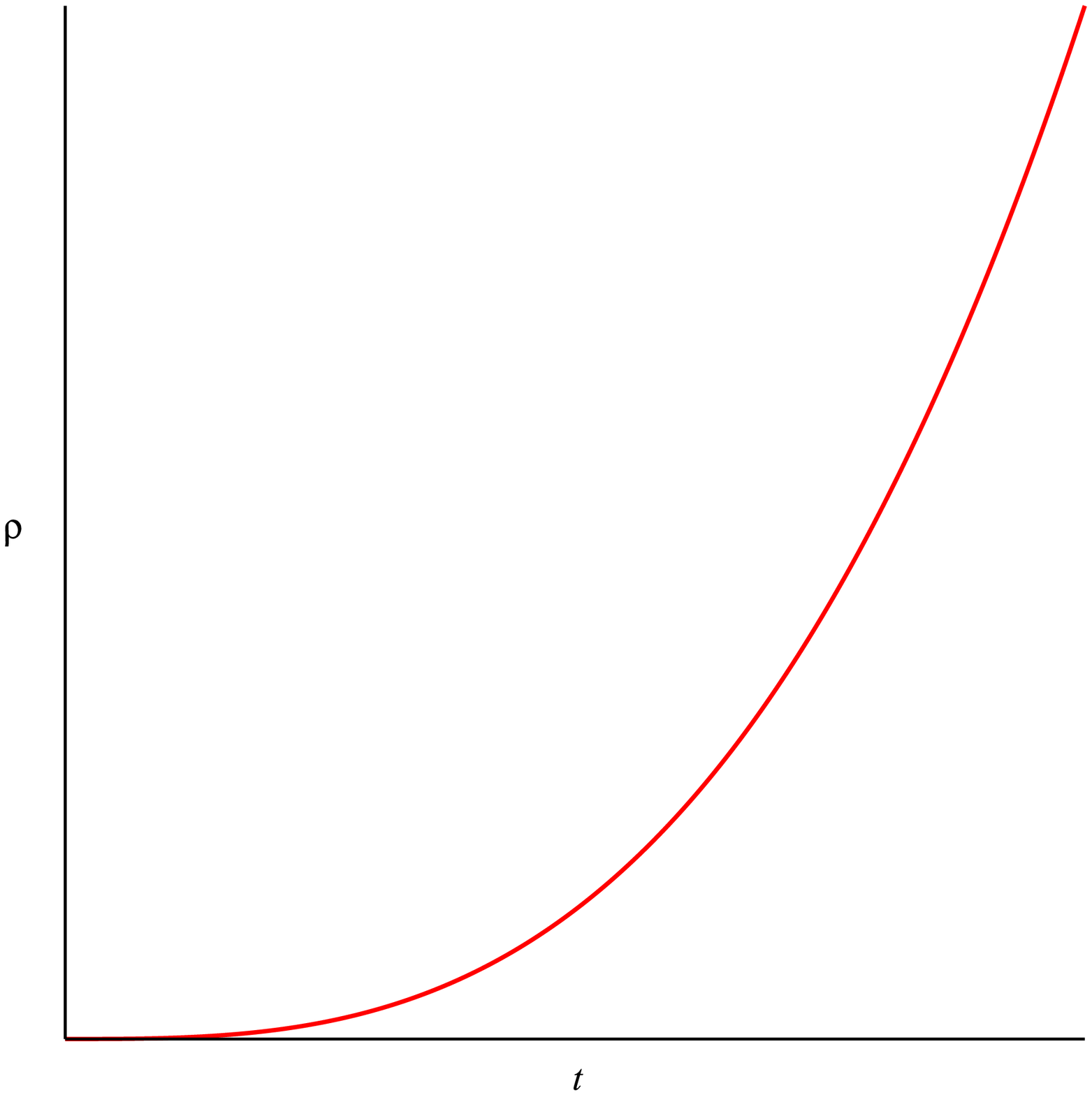} \vspace{9cm}\includegraphics{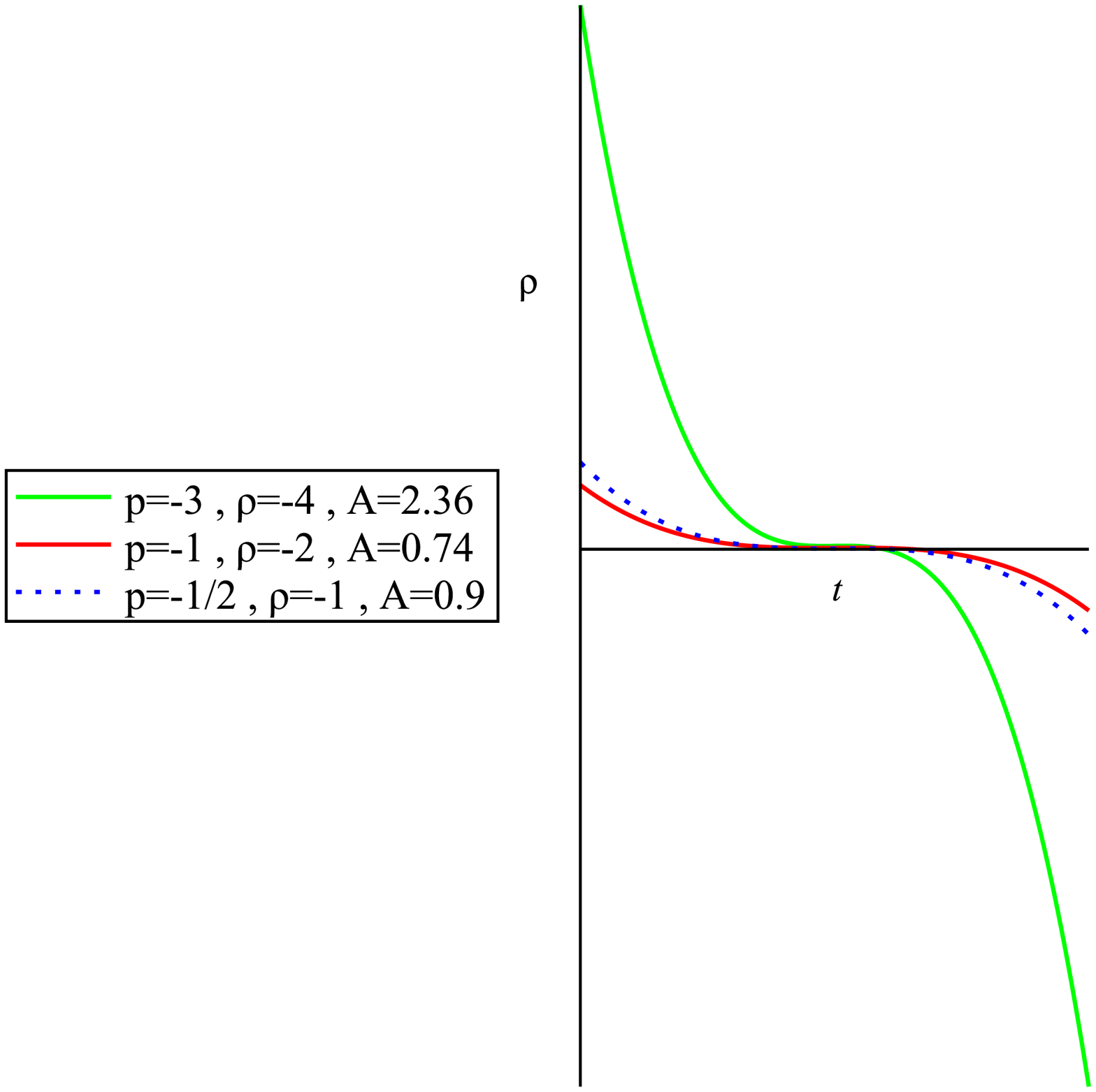}
\vspace{10cm}\includegraphics{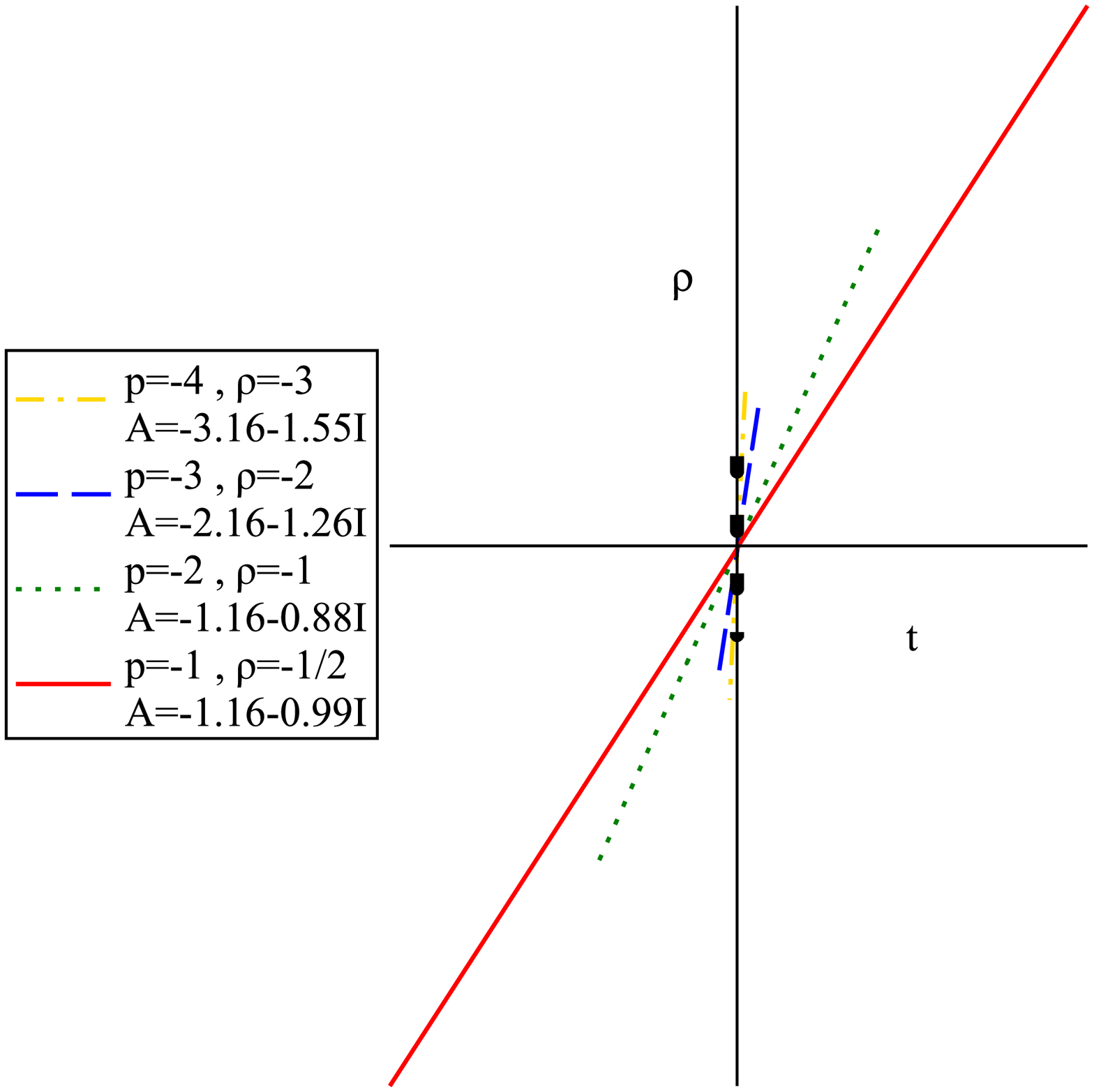}
\end{center}
\caption{\small {The effects of energy density in term of cosmic
time on the universe expansion in Israel-Stewart theory(IS).
Fig.5.A(up): standard approach with condition $p=-\rho$ and
$A=\frac{1}{3}\mp\frac{1}{6\sqrt{\rho}}$, Fig.5.B(middle): dynamical
and non-phantom approach with condition $p>-\rho$ and
$A>\frac{1}{3}\mp\frac{1}{6\sqrt{\rho}}$, Fig.5.C(down): phantom
approach with condition $p<-\rho$ and
$A<\frac{1}{3}\mp\frac{1}{6\sqrt{\rho}}$. In this mode we considered
the value of $t_{\alpha}=10$ and value of the Hubble constant
$H\simeq71\times10^{9}$}.}
\end{figure}
\begin{table}
\begin{center}
\caption{The results of the energy density effects in terms of
cosmic time on the universe expansion in the framework of IS theory
({\bf Mode 2}).} \vspace{0.5 cm}
\begin{tabular}{|p{3cm}|p{4cm}|p{7.7cm}|}
  \hline
  \hline {\tiny Condition} & {\tiny Approach} & {\tiny Description} \\
  \hline {\tiny $P=-\rho \newline A=\frac{1}{3}\mp \frac{1}{6\sqrt{\rho}}$}& {\tiny Standard }& {\tiny In this case, it is the same as in Table (1), except that the energy density is measured in terms of scale factor, but here it is defined by cosmic time and also a constant A. With Over time, the universe has an accelerated expansion (of course, before time $t_{\alpha}$). After time $t_{\alpha}$, we assume that over time pressure tends to $p\longrightarrow - 2\rho$ and the universe expand slowly. In this approach the energy density value
  increases rapidly but its value is symmetrical with the value of negative pressure.}\\
  \hline {\tiny $P>-\rho\newline A>\frac{1}{3}\mp \frac{1}{6\sqrt{\rho}}$}& {\tiny Dynamical and non-phantom }& {\tiny In this case, with over time, the pressure value is negative and will change with the energy density value[20], and expansion will increase, and as a result, the energy density will decrease (according to condition A). Therefore, the condition for the expansion of the universe with over time is the decrease in the dark energy density variable value. As much as the energy density value is lower, the universe expand faster.}\\
  \hline {\tiny $P<-\rho\newline A<\frac{1}{3}\mp \frac{1}{6\sqrt{\rho}}$} &{\tiny Phantom} &{\tiny In this case, with over time, the energy density becomes bigger and tends to be infinity. As a result, the universe expands more rapidly, but after crossing the phantom divide line [21], the value of energy density becomes infinity(the maximum value of energy density), the universe's expansion rate gradually decreases and finally reaches a big rip singularity after time $t_{\alpha}$.}\\
  \hline
\end{tabular}
\end{center}
\end{table}
\section{Summary}
In accordance with the unknown nature of dark energy, some dark
energy models has been studied as a cosmic fluid in the framework of
thermodynamic laws. In this regard, viscosity is a feature of the
universe fluid content as discussed in the present research.
Therefore, in the first part of this article, we expressed the
thermodynamics of cosmic fluids in general with a constant and
variable equation of state under the two theories of Eckart and
Israel-Stewart.  Then, we examined the dissipative effects of cosmic
fluids and finally examine the effects of energy density for dark
energy in the Israel-Stewart(IS) theory. The results are organized as follows:\\
We first investigated the thermodynamics of cosmic fluids in the
dark energy bulk viscosity model and the general relationships.
Then, we expressed the thermodynamic relationships of Eckart's
theory. In IS theory, a differential equation is defined in terms of
temperature $(T)$, bulk viscosity coefficient $(\xi)$ and relaxation
time $(\tau)$, and each of these parameters is obtained in terms of
energy density. First time, we obtained an equation for the bulk
viscosity pressure by placing the value of $s=1$ in equation
$(37)$ and the Ansats solution.\\
In the third section, we investigated the dissipative effects of
cosmic fluids and we concluded the best theory for studying the
effects of dissipation is IS theory, because it develops dissipation
processes with slower rate. In this section, We presented two
scenarios. In the first scenario, we considered the cosmic fluid to
be non-adiabatic. We concluded that this type of fluid conforms to
the standard model of cosmology and cosmic observations, which
during it, entropy and temperature increase and causes the universe
to expand. But in the second scenario, we considered the cosmic
fluid to be adiabatic and concluded that no entropy is produced in
this scenario and temperature decreases. Also, we defined two states
in this scenario, one is the accelerated expansion state and the
other is a phantom state that leads to a singularity at the late of
the universe. In the continuation of the third section, we
investigated the effects of cosmic fluid entropy in general and then
in the framework of Eckart and IS theories. We concluded that IS
theory better represents the production of entropy in dissipated
cosmic fluids. Also, in this part we defined two different scenarios
for entropy. The first scenario represents more correct behavior
from cosmic fluid entropy with over time but the second scenario is
completely opposite to the first scenario.\\
In the fourth section, first time we expressed the effects of energy
density on the expansion of the universe in the framework of IS
theory. According to the definition of bulk viscosity pressure in IS
theory, we suggested two modes, one is energy density in terms of
scale factor and the other is energy density in terms of cosmic time
and an expansion descriptor coefficient $(A)$. The obtained results
in this section are comprehensively presented in two tables $(1)$
and $(2)$. We also plotted the energy density behavior in three
standard, dynamical and phantom approaches and obtained its
results.\\\\
{\bf Acknowledgements} We would like to thank all those who helped
us gather the information needed for this research.\\
{\bf Funding} The authors did not receive support from any
organization for the submitted work.\\
{\bf Data availability} The data that support the findings of this
study are available on request from the corre- sponding author.\\
{\bf Conflict of interest} The authors declare that they have no
known competing financial interests or personal relationships that
could have appeared to influence the work reported in this paper.
This research did not receive any specific grant from funding
agencies in the public, commercial, or not-for-profit sectors.\\
{\bf Ethics approval} We have followed all the rules stated in the
Ethical Responsibilities of Authors section of the journal.\\
{\bf Consent for publication} We consent to the publication of this
manuscript with the rules of the journal.\\

\section{Appendix}
\setcounter{equation}{0} In this section, we consider some equations
that were involved in obtaining the equations in the previous
sections. To calculate entropy we need to calculate the particles
number density of the fluid and the temperature of the fluid.
According to equation $(33)$ and its solution
$(T=T_0\rho^\frac{p}{p+\rho})$ we can write:
\begin{equation}\dot{T}=(\frac{p}{p+\rho})T\frac{\dot{\rho}}{\rho}+\rho^\frac{p}{p+\rho}\frac{df}{dx}\dot{x}\end{equation}
where $df$ is a function in term of $x=T_0\rho^\frac{p}{p+\rho}$ and
also according to equations $(3)$ and $(32)$ we can write for
$\dot{x}$:
\begin{equation}\dot{x}=\frac{\rho^\frac{p}{p+\rho}}{n}[(\frac{\rho}{p+\rho})\frac{\dot{\rho}}{\rho}-\frac{\dot{n}}{n}]=0 \end{equation}
then appendix equation (1) becomes equation
$\frac{dT}{T}=-3\frac{p+\rho}{\rho}\frac{da}{a}=\frac{p}{p+\rho}\frac{d\rho}{\rho}$
in the bulk viscosity model of a fluid with coefficient $\xi$ that
$\xi$ is constant to calculate Hubble parameter and scale factor, we
have(which equations $(74)$ and $(75)$ are concluded from them)
[18]:
\begin{equation}H(t)=\frac{H_0\xi_0e^\frac{3\xi_0 (t-t_0)}{2}}{H_0(\frac{p+\rho}{\rho})(e^\frac{3\xi_0 (t-t_0)}{2}-1)+\xi_0}\end{equation}
\begin{equation}a(t)=(1+\frac{H_0(p+\rho)}{\xi_0\rho}(e^\frac{3\xi_0 (t-t_0)}{2}-1))^\frac{2\rho}{3(p+\rho)}\end{equation}
which by integrating the Hubble parameter relation, the scale factor
is obtained with condition $a(t_0)=1$.\\
Finally, the energy density of the fluid can be obtained in term of
scale factor:
\begin{equation}\rho(a)=\rho_0(\frac{\xi_0\rho}{H_0(p+\rho)}+(\frac{H_0(p+\rho)-\xi_0\rho}{H_0(p+\rho)})a^\frac{-2\rho}{3(p+\rho)})^2 \end{equation}
Now, if $\xi$ is assumed to be variable as
power-law,$\xi=\alpha\rho^s$,we will have [18]:
\begin{equation}H(t)=H_0(1+\frac{3H_0}{2}(\frac{p+\rho}{\rho}-\alpha)(t-t_0))^{-1}\end{equation}
\begin{equation}a(t)=(1+\frac{3H_0}{2}(\frac{p+\rho}{\rho}-\alpha(t-t_0)))^\frac{2}{3(\frac{p+\rho}{\rho}-\alpha)}\end{equation}
\begin{equation}\rho(a)=\rho_0a^{-3(\frac{p+\rho}{\rho}-\alpha)}\end{equation}
To calculate the fluid temperature in the bulk viscosity model in
Israel-Stewart(IS) theory in term of scale factor according to the
relations $(41)$, $(44)$ and $T=T_0\rho^\frac{p}{p+\rho}$ we will
have:[16]
\begin{equation}T(t)=\tilde{T}(3A^2)^\frac{p}{p+\rho}(t_\alpha-t)^\frac{-2p}{p+\rho}=
\tilde{T}(3A^2(t_\alpha-t)^{-2})^\frac{p}{p+\rho}a^\frac{2p}{A(p+\rho)}\end{equation}

\end{document}